\documentclass{egpubl}
\usepackage{pg2026}

\usepackage{amsmath}

\usepackage{graphicx}

\usepackage{xspace}
\usepackage{microtype}
\usepackage{wrapfig}
\usepackage{siunitx}

\usepackage{amsfonts}
\newcommand{\romy}[1]{{\color{blue} \small(Romy: #1)}}

\newcommand{\update}[1]{{ #1}}

\newcommand{\name}{Blended Chart Surfaces\xspace}
\newcommand{\localmaps}{local maps\xspace}

\usepackage{enumitem}
\usepackage{array}

\usepackage[normalem]{ulem} 
\usepackage{xcolor}         
\usepackage[table]{xcolor}

\usepackage{tikz}

\SpecialIssuePaper         

\CGFStandardLicense

\usepackage[T1]{fontenc}
\usepackage{dfadobe}  

\usepackage{cite}  
\BibtexOrBiblatex
\electronicVersion
\PrintedOrElectronic

\usepackage{egweblnk}

\usepackage{hyperref}
\usepackage{cleveref}

\title[Blended Chart Surfaces]%
      {Blended Chart Surfaces: A Seamless Explicit Representation for
Smooth Surface Fitting
\vspace*{-.4in}
}

\author[R. Williamson \& N. J. Mitra]
       {Romy Williamson$^{1}$\orcid{0009-0000-9833-7821}
        and Niloy J. Mitra$^{1,2}$\orcid{0000-0002-2597-0914}
        \\
        $^1$University College London, 
        $^2$Adobe Research London\\
        \vspace*{-.3in}
       }

\begin{document}


\maketitle
\begin{abstract}


A surface representation suitable for geometry processing should be \emph{compact} and \emph{explicit}, provide \emph{global smoothness guarantees}, support a wide range of surface \emph{topologies}, and offer reliable access to \emph{differential quantities} such as normals and surface energies, while remaining compatible with modern differentiable optimization.
Yet existing neural representations typically sacrifice one or more of these properties: implicit fields typically require iso-surfacing for downstream use, while explicit neural maps are constrained by canonical-domain parametrizations and/or exhibit seam artifacts between local charts.
We introduce \emph{Blended Chart Surfaces (BCS)}, a compact, network-free, explicit representation that is \emph{smooth by construction} and anchored to user-provided topology.
Given a coarse proxy mesh encoding the intended surface topology and approximate geometry, \name jointly optimize for a polynomial map at each proxy vertex using an off-the-shelf optimizer to fit to an implicit target shape, avoiding the need for an input parametrization. Neighboring maps are fused using a smooth `one-ring coordinate' blending scheme, decoupling surface topology and coarse geometry (carried by the proxy) from geometric details (carried by the smooth local patches).
The resulting surface is globally smooth, fully differentiable, and enables stable evaluation of positions and derivatives, making differential quantities and surface energies directly accessible.
Additionally, our construction is equivariant to rigid motions and scaling of the proxy mesh.
We evaluate \name on surfaces spanning varying topology and geometric complexity, and compare against explicit alternatives including interpolating-function baselines and mesh-displacement MLPs. 
Across these, \name achieves a favorable trade-off among compactness, simplicity, access to differential quantities, and expressivity while remaining smooth across patch boundaries.
\textit{\href{https://geometry.cs.ucl.ac.uk/projects/2026/bcs/index.html}{Link to project webpage.} }

\end{abstract} 


\section{Introduction}

A surface representation suitable for modern geometry processing should be compact, expressive, support a wide range of object topologies and geometric details, and provide smoothness guarantees (e.g., $C^k$ or $C^\infty$) and access to differential quantities, including normals and surface energies. At the same time, it should be compatible with modern learning pipelines: differentiable, compatible with optimization and fitting, and amenable to end-to-end training.

\begin{figure}[t!]
  \centering
  \includegraphics[width=.9\columnwidth]{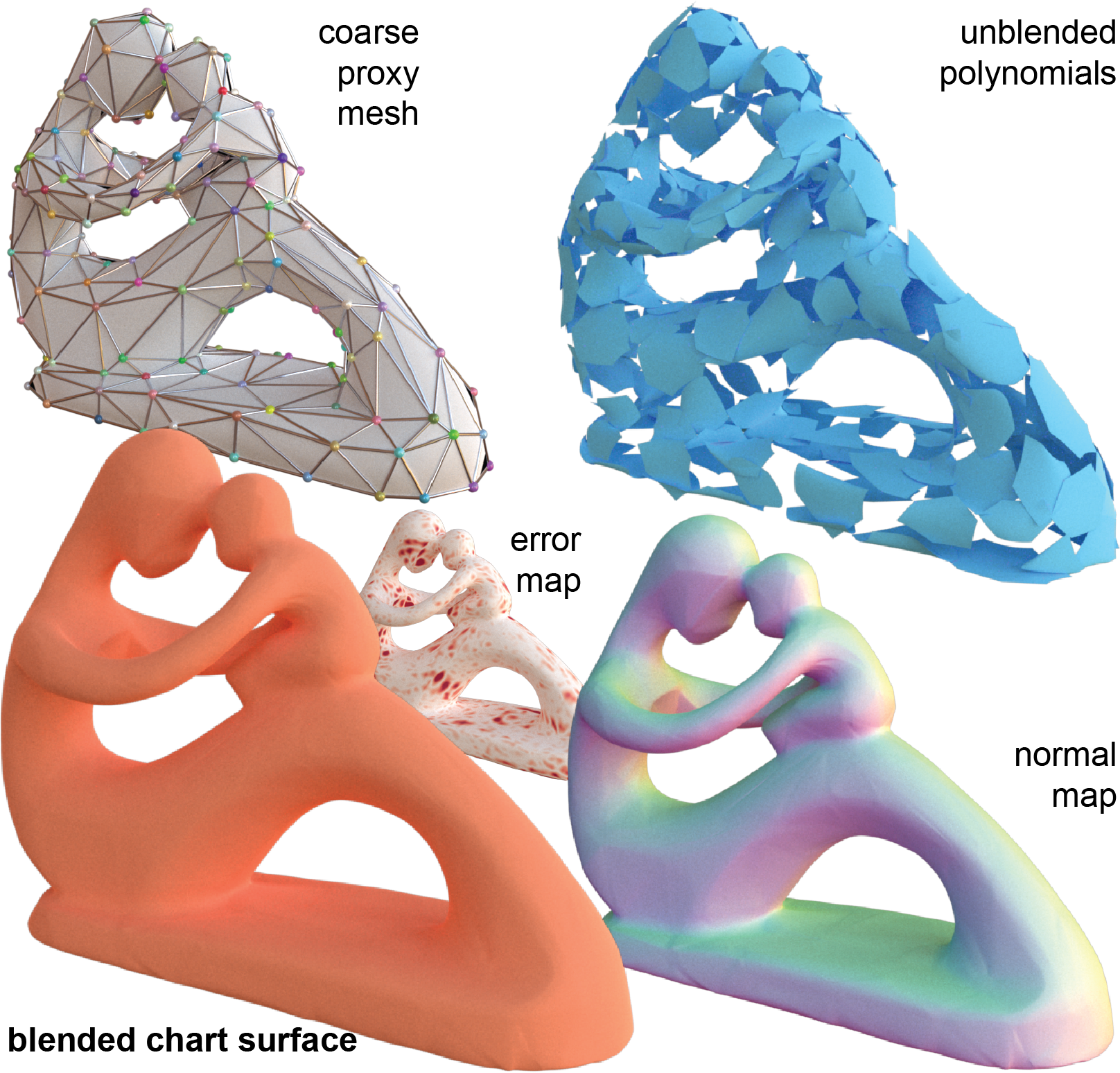}
  \caption{
    We propose \name, a coarse proxy mesh--guided, network-free, explicit surface representation formed by composing local polynomial maps. 
    The model is compact, 
    faithfully captures surface geometry, and is fully differentiable for optimization in modern learning pipelines. 
    Here, $|V|=250$ and $d_\mathrm{poly}=2$. 
  }
  \label{fig:teaser}
  \vspace*{-.2in}
\end{figure}

\begin{figure}[t!]
    \centering
\includegraphics[width=\columnwidth]{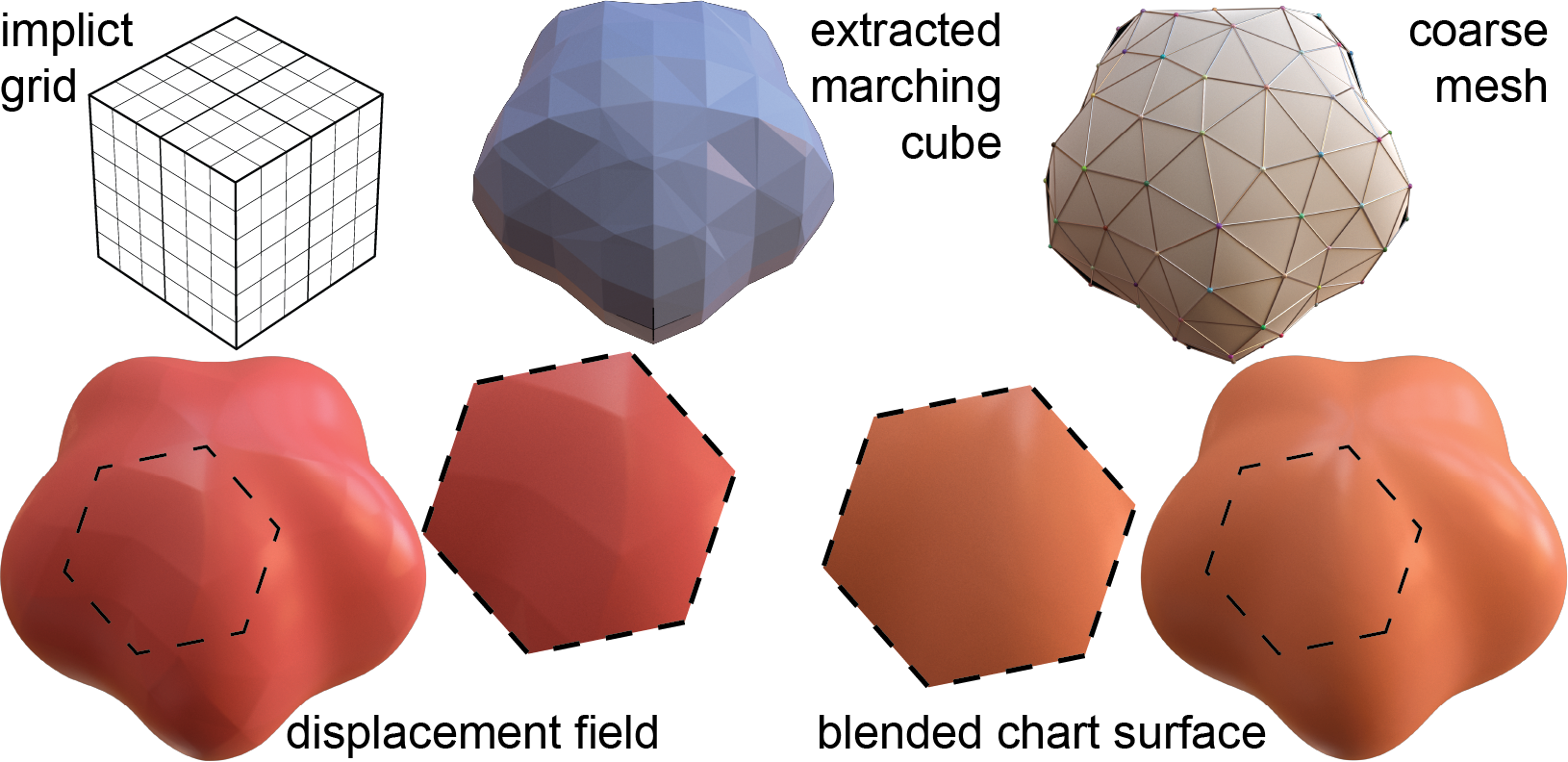}
\caption{\textbf{Motivation.} (Top-left) Implicit fields (stored on a grid or as a neural field) require high effective resolution (here iso-surfaced from a $10^3$ grid) to avoid visible discretization artifacts when extracted with Marching Cubes. (Bottom-left) Mesh-based displacement fields parametrized by an MLP (e.g.~\cite{NeuralGeomFields}) require fewer parameters but retain the tangent discontinuities present in the proxy. (Right) Our \name learns on a coarse proxy mesh ($|V|=107$ vertices here) yet yields an explicit surface that is, by construction, $C^\infty$ smooth across patch boundaries.}
    \label{fig:motivation}
\end{figure}

Implicit representations (e.g., occupancy fields~\cite{mescheder2019occupancy}, signed distance fields~\cite{ohtake03,park2019deepsdf,ErlerEtAl:Points2Surf:ECCV:2020,zhang23}) are a natural fit for both classical and learning-based pipelines: they are easy to define, easy to optimize, and  handle complex geometries and topologies in a unified setting. However, they are inherently volumetric fields ($\mathbb{R}^3\rightarrow\mathbb{R}$) and therefore not as compact as surface-based models. Moreover, downstream applications (e.g., engineering, manufacturing) typically require an explicit surface for sampling, collision handling, or simulation, which is commonly obtained by iso-surfacing the implicit fields via Marching Cubes~\cite{marching-cubes1987}, Dual Contouring~\cite{dual-contouring2002}, or related methods. Such conversions produce piecewise-linear approximations that can introduce non-smooth artifacts and degrade the stability of differential estimates, effectively breaking the smoothness of the underlying surface.

Explicit (surface) representations ($\mathbb{R}^2\rightarrow\mathbb{R}^3$), in contrast, avoid the costs associated with volumetric representations and provide direct access to surface evaluation and differential quantities; existing options, however, impose restrictive trade-offs. Classical smooth representations such as splines~\cite{piegl1997nurbs}, subdivision surfaces~\cite{warren2002subdivision}, polyhedral-net splines~\cite{peters2023}, isogeometric surfaces~\cite{cottrell2009isogeometric} offer continuity guarantees; 
while they are designed to interpolate any coarse net, they are not suitable for fitting to a target surface in a learning based setup. At the other end of the spectrum, dense triangle meshes remain prevalent in practice, but often require hundreds of thousands of faces or more to capture surface detail; further, they necessitate special discrete differential geometric estimates. Neural explicit alternatives provide differentiable parametrizations~\cite{atlasNet18,morreale2021neural,williamson2025spherical,Li-2025-MASH}, but many are restricted by canonical-domain topology assumptions (e.g., disc or sphere topologies), while patch-based variants suffer from seams, derivative discontinuities~\cite{NeuralGeomFields}, or imperfect stitching across chart boundaries~\cite{betterpatch2020,atlasNet18}.

This motivates a natural question: can we obtain an \emph{explicit} surface representation that remains \emph{compact} and \emph{expressive}, while also providing \emph{smoothness guarantees} and being compatible with modern differentiable optimization?
We propose \textit{\name}, which builds a globally smooth surface on top of a user-provided coarse simplicial proxy mesh that specifies the intended object topology and coarse geometry. We assign a local polynomial map to each proxy vertex and blend neighboring maps using partition-of-unity blending, decoupling topology (carried by the proxy mesh) from geometry (carried by the smooth local patches).
Technically, we enable this by a novel `one-ring coordinate' system (see  \cite{interpolatingSplines} for an alternate formulation) and enumerating the properties required for suitable blending functions. 
Our representation is network-free, explicit, compact, and smooth by construction; most importantly, it remains fully differentiable, enabling direct optimization of patch coefficients against targets such as implicit fields, point samples, or analytical surfaces, without requiring access to any global surface parametrization as input. For example, the \texttt{fertility} model in \Cref{fig:teaser} uses quadratic patches (18 scalar coefficients per vertex) at each of the $250$ vertices of the coarse base mesh, and the patches are `blended' to represent the full surface. More generally, our representation requires  $\mathcal{O}(d_\text{poly}^2|V|) $ parameters, where $|V|$ denotes the number of vertices in the proxy mesh and $d_{\mathrm{poly}}$ the polynomial degree (e.g. $d_\text{poly}=2$ for quadratic patches).

We evaluate \name on shapes of varying topology and geometric complexity, and compare against existing explicit surface representations supporting surface fitting, including several families of interpolating functions over the same proxies and mesh-based displacement functions (similar to Neural Geometry Fields for Meshes~\cite{NeuralGeomFields}). 
Our results show that \name achieve a favorable trade-off between compactness and expressivity while remaining smooth across patch boundaries by construction, and that they can be optimized effectively within modern learning and differentiable optimization pipelines.
In summary, we:\\
(i)~introduce a compact, explicit surface representation as a set of overlapping local polynomial patches blended with partition-of-unity weights, yielding global smoothness and enabling end-to-end differentiation and surface fitting; \\
%
(ii)~encode a target surface on a given coarse mesh by optimizing patch coefficients directly, without expecting any global charting as input while supporting a broad range of object topologies specified by the proxy, naturally accommodating difficult topologies such as \emph{open boundaries} and \emph{non-orientable} surfaces;   \\ 
(iii)~admit progressive levels of detail by increasing polynomial degree (e.g., constant, linear, quadratic, cubic) and/or refining the proxy, providing a continuum between compactness and faithful reconstructions (i.e., surface fitting). 

\section{Related Work}

\paragraph*{Classical surface representations.}
Implicit representations (e.g., signed distance functions) are expressive and naturally accommodate complex geometry and topology; earlier methods focused on extracting such implicit fields from sparse observations and/or point sets~\cite{ohtake03,poissonSurfRec,kolluri08}. In practice, however, they represent a \emph{volumetric} field over $\mathbb{R}^3$ and are therefore inefficient as surface models when downstream applications require a 2D manifold embedded in 3D; moreover, many geometry-processing and simulation tasks require an explicit surface for sampling, collision handling, and numerical discretization, typically obtained via iso-surfacing such as marching cubes~\cite{marching-cubes1987} or dual contouring~\cite{dual-contouring2002}, which yields piecewise-linear (or low-order) approximations that can introduce artifacts and degrade the smoothness and stability of differential estimates. 

Explicit representations, in contrast, are easy to sample from and provide direct access to surface evaluation, differential quantities, and surface energies, forming the backbone of polygonal geometry processing. Spline and subdivision surfaces offer strong continuity guarantees and interpretability, but complex shapes often require multiple patches and enforcing $C^k$ continuity across boundaries is nontrivial; in many practical settings, global guarantees depend on structural restrictions on the control net. Notably, Grimm and Hughes~\shortcite{grimm95} present an atlas-inspired spline construction using overlapping basis functions to achieve global $C^k$ continuity, though the formulation is complex due to multiple chart types and valence restrictions. Subsequently, Ying and Zorin~\cite{ying2004} provide a more principled and simpler manifold-based construction that achieves $C^\infty$ continuity on arbitrary topologies by utilizing a partition of unity weighting over a set of overlapping charts. \update{Gu et al. ~\cite{ManifoldSplinesGu05} propose a different atlas-based formulation, using `affine' atlases.}
Despite their theoretical elegance, these methods have not yet been widely adopted for surface optimization and fitting in the context of deep learning.

A closely related direction for constructing smooth surfaces from arbitrary control meshes is the framework of polyhedral-net splines proposed by Peters and colleagues~\cite{peters1994smooth,peters2023}. These methods define a systematic way to generate $C^1$ or $C^2$ surfaces by mapping the faces of an irregular polyhedral net to specialized spline patches, effectively generalizing box splines to arbitrary topology. While these constructions provide a robust bridge between polygonal control structures and smooth analytic surfaces, they typically rely on complex, localized stencil definitions to handle extraordinary vertices, which makes them cumbersome to implement for arbitrary-order continuity or within modern differentiable geometry pipelines.

A parallel line of research focuses on generalized blending schemes and $n$-sided patches to overcome topological constraints. Fang~\cite{fang2023} proposes a generalized blending scheme for arbitrary orders of continuity on triangular and polygonal meshes, while Salvi~\cite{salvi2026} utilizes blended $n$-sided interpolants for polyhedral design on quad meshes. These blending philosophies are also utilized for curve design; for instance, Jiang and Chen~\cite{jiang2025} developed $G^2$ interpolating splines with local maximum curvature control, offering favorable properties for high-quality boundary construction.

To bridge the gap between simple meshes and complex splines, other approaches utilize specialized elements or generalized coordinates. Anisimov et al.~\cite{anisimov2017} introduced blended barycentric coordinates, allowing for smooth interpolation over arbitrary planar polygons. Similarly, Powell-Sabin elements have been employed to achieve low-order smoothness by averaging first-order vertex information, as seen in the construction of algebraic smooth occluding contours~\cite{capouellez2023}. Alternatively, specialized data-driven subdivision stencils~\cite{neuralSubdivision:20} have been proposed to replace classical rules, although they often retain assumptions about the underlying control mesh. Despite these advances, polygonal meshes remain ubiquitous due to their simplicity; however, capturing fine detail and obtaining stable differential estimates often require high mesh resolution (e.g., $50k+$ faces), and piecewise-linear discretizations introduce sharp edges by construction, complicating robust definitions of curvature and higher-order surface operators~\cite{ddg:06:course, ddg:13:course}.

The most closely related to our surface construction is the recent work of Djuren et al.~\cite{interpolatingSplines}, who propose interpolating splines over triangulated surfaces by blending vertex-centric local geometries. While their method shares our core philosophy of utilizing one-ring neighborhoods for surface construction, their approach is primarily focused on geometric interpolation and does not address the challenges of surface fitting. In contrast, our work is specifically designed for differentiable geometry processing, enabling the use of such representations in surface learning and optimization tasks. Djuren et al. choose vertex functions \textit{independently} and force every vertex function to interpolate its one-ring vertices, which is an over-constraint, because the vertex functions are subsequently blended and do not necessarily need to be close before the blending occurs. Our observation is that end-to-end optimization through the blending function is preferable for accurate reconstruction: our Blended Chart Surface construction uses all its degrees of freedom for improving surface reconstruction (working with, not against, the blending function) and the `unblended' patches are not forced to satisfy unnecessary constraints. We provide a comparison with the polynomial version of the Interpolating Spline construction in our evaluation (see Section~\ref{sec:evaluation}). In terms of the construction, their `conformal flattening' retains the relative size of angles in a one-ring, but their blending construction is only $C^2$ continuous, whereas we guarantee $C^\infty$ continuity but do not maintain relative angle sizes.

\paragraph*{Function approximation in machine learning.}
Classical approaches to function approximation include Fourier representations and basis interpolation methods such as radial basis functions~\cite{saupe20013d,iglesias2004functional,GORDON197495}. More recently, neural approaches have emphasized \emph{overfitting} a single signal (e.g., images, volumes, radiance, materials, etc.) with a highly expressive function class, enabling accurate reconstruction and convenient differentiation. Two families of constructions are especially common for encoding fields: (i) direct mapping~\cite{sitzmann20,tancik2020fourfeat,nerf:21} with an MLP from coordinates to values, which can be expressive but computationally expensive and difficult to interpret; and (ii) hybrid schemes~\cite{muller22,ReluField_sigg_22} that store features on a grid (or multi-resolution grids) and use interpolation of nearby features (e.g., linear interpolation~\cite{takikawa2021nglod}, radial basis function~\cite{carr01} interpolation, etc.), followed by lightweight MLPs, which have local impact (i.e., based on chosen neighborhood definition) and are computationally efficient.


\paragraph*{Neural implicit volumetric fields.}
Neural implicit fields represent surfaces via occupancy or signed distance values predicted by a neural function ($\mathbb{R}^3\rightarrow\mathbb{R}$), a special case of function approximation discussed above. They are widely used in reconstruction and generative modeling, with prominent examples including DeepSDF~\cite{park2019deepsdf}, IM-Net~\cite{Chen_2019_CVPR}, Occupancy Networks~\cite{mescheder2019occupancy}, and Shape2Vec~\cite{zhang23}. 
These representations integrate naturally with modern optimization and learning pipelines, but they remain volumetric functions and therefore inherit the same downstream limitation as classical implicits.

%

\paragraph*{Neural explicit surfaces.}
A notable approach is patch-based modeling~\cite{atlasNet18}, where multiple local parametrizations are learned and stitched together to represent a surface as a collection of patches obtained by mapping planar domains ($\mathbb{R}^2\rightarrow\mathbb{R}^3$). While it avoids requiring a global parametrization, it provides no continuity guarantees across patch boundaries, and generated surfaces can contain gaps or overlaps; subsequent work improves stitching but still highlights the challenges of obtaining seamless reconstructions from independently parametrized patches~\cite{betterpatch2020}. Recently, Neural Geometry Fields for Meshes (NGFM)~\cite{NeuralGeomFields} also uses a patch-based formulation over a coarse quadrilateral scaffold: local patch displacement functions are defined from per-vertex quantities via bilinear interpolation, which enforces $C^0$ continuity but does not guarantee higher-order continuity across patch seams (e.g., along edges and at vertices); see \Cref{fig:motivation}. 

Other methods assume access to global parametrizations by mapping a single canonical domain to the surface. Neural Surface Maps~\cite{morreale2021neural} represent a surface as the image of a single disc-like domain, concentrating discontinuities at a cut but restricting the class of representable topologies. Spherical Neural Surfaces~\cite{williamson2025spherical} instead use a spherical domain to avoid seams entirely, yielding a seamless explicit representation, but the approach is limited to genus-$0$ surfaces (homeomorphic to the sphere).
Similarly, NeuPPS~\cite{yang2024neupps} represents surfaces as neural piecewise parametric patches to capture high-frequency details, yet, like its predecessors, ensuring rigorous global smoothness and topological consistency across the entire collection of patches remains a significant hurdle. 
In contrast, in our method we avoid canonical-domain topology restrictions by anchoring the representation to a user-provided proxy mesh, and avoid seam artifacts by enforcing $C^\infty$ smoothness across overlaps, via a patch-blending formulation.

\section{Method}

\paragraph*{Overview.}
Our goal is to construct an explicit representation on top of a user-provided coarse proxy (a polyline for curves, or a triangle mesh for surfaces) such that the resulting curve/surface has guaranteed continuity properties (smoothness).
The key challenge is that the proxy itself is generally non-smooth, containing corners or sharp edges, but the final Blended Chart Surface (or Curve) should be smooth. 
Our approach assigns \emph{vertex functions} (we used low-degree polynomials) to proxy vertices and combines neighboring vertex functions using suitably designed \textit{transition maps} and \emph{blending functions} that---by construction---enforce consistency and smoothness across overlaps. \update{Note that the type of smoothness that is guaranteed is `function' continuity rather than `geometric' continuity. This means that the local parametrization function (taking points from $\mathbb{R}^2$ to $\mathbb{R}^3$) is guaranteed to be $C^\infty$ continuous, but does not \textit{by construction} rule out the possibility of geometric/visual non-smoothness if the Jacobian of the map were exactly zero at some point. Please see Supplemental B.9 for further discussion.}
Note that the given proxy specifies the curve/surface topology and coarse geometry, while the detailed geometry is obtained by directly optimizing the parameters of the vertex functions, using Adam~\cite{adam}, against a target implicit representation (e.g., an analytical SDF, or deep implicit field), without requiring a precomputed parametrization of the target curve or surface.

\subsection{Blended Chart Curves (1D)}
\label{subsec:BPC}

\begin{figure}[t!]
    \centering
    \includegraphics[width=\columnwidth]{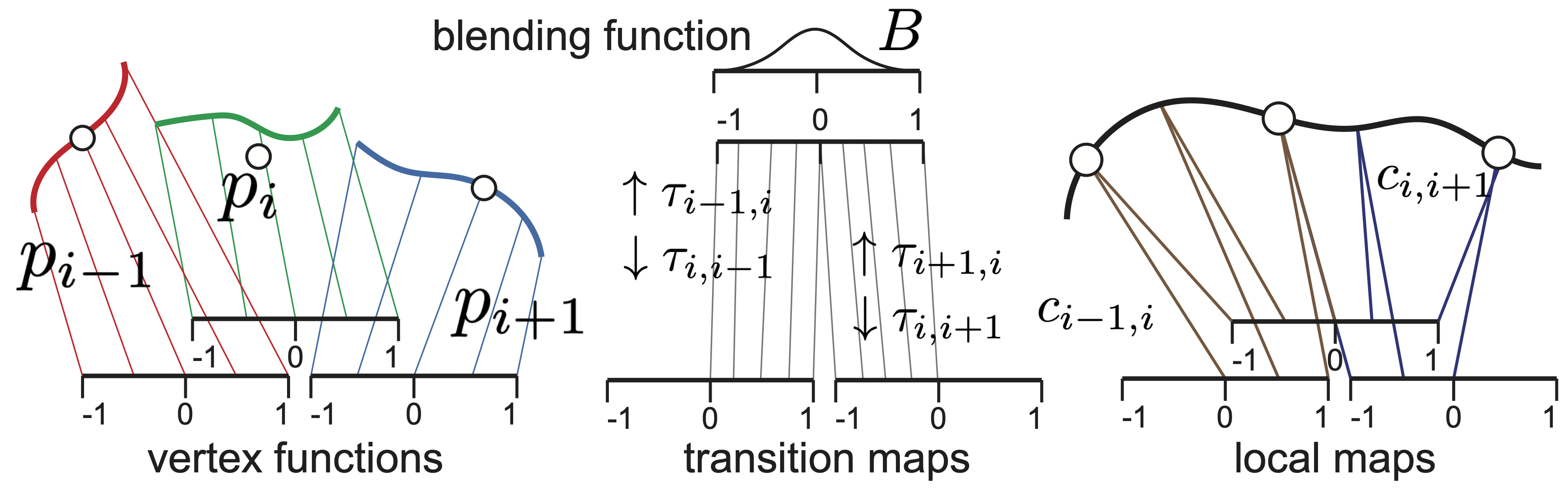}
    \caption{\textbf{Curve case.} Each coarse proxy vertex is assigned a polynomial \emph{vertex function}. Within each edge, the vertex functions are combined using \emph{blending functions} (partition-of-unity weights), producing a \emph{local map} that evaluates curve points (and derivatives) smoothly on that base domain. See text for details. }
    \label{fig:1d_blending}
\end{figure}

Let $(V,E)$ be a coarse manifold closed curve (no T-junctions or self-intersections), given as a closed polyline with $n = |V|$ vertices and $n$ edges.
Our goal is to construct a smooth curve on top of this discrete control curve by assigning a \emph{vertex function} $p_i$ to each coarse vertex $v_i \in V$, and \emph{blending} adjacent vertex functions according to the connectivity specified in $E$.
Without loss of generality, we order the vertices such that 
$
E := \{ (0,1), (1,2), \ldots, (n-1,0) \};
$
thus, vertices are adjacent exactly when they have consecutive indices modulo $n$. In what follows, all indices are modulo $n$. See \Cref{fig:1d_blending}. 

\paragraph*{Overlapping domains and transition maps.}
We treat the domains of adjacent local maps as overlapping copies of $[-1,1]$, where the second half of one interval is identified with the first half of the next. In atlas terminology (see Chapter 1 of~\cite{leeIntroSmoothManifolds}), the transition maps between neighboring domains are, 
\begin{equation}
\begin{matrix}
    \tau_{i,i+1} : [0,1] \rightarrow [-1,0],\quad t \mapsto t - 1,\\
    \tau_{i+1,i} : [-1,0] \rightarrow [0,1],\quad t \mapsto t + 1.
\end{matrix}
\end{equation}
This means that if $t$ represents a position in the coordinates of domain $A$, in a part of the domain that overlaps with domain $B$, then $\tau_{A,B}(t)$ represents the same position in the coordinates of domain $B$. Although these transition maps are simply unit translations, we have kept the $\tau$ notation in order to highlight the `atlas' perspective.

\paragraph*{Consistency.}  
A collection of \textit{local maps} $\{c_i\}_{i=0}^{n-1}$ defines a single continuous curve if the maps agree on overlaps, i.e.,
\begin{equation}
c_\alpha \equiv c_\beta \circ \tau_{\alpha,\beta}
\quad \text{on the domain of}\ \tau_{\alpha,\beta},
\label{equation:consistency}
\end{equation}
for any pair with overlapping domains.
If the $c_i$ are homeomorphisms, then their inverses $c_i^{-1}$ form charts of an atlas. Note that our construction is not necessarily a true atlas because we do \emph{not} constrain the $c_i$ to be homeomorphisms (they may not be invertible). In these cases the curve could contain cusps, although this is not a practical issue because the set of parameter-values producing non-invertible $c_i$ maps has measure zero (see supplemental section B.9), and it does not affect the computation of derivatives. Our construction is an example of the broad class of surfaces that Gallier et al. \cite{gallier12} refer to as a `Parametric Pseudo-Manifolds' (although the construction methods are different).

\paragraph*{Blending vertex functions.}
We construct the local maps $c_i$ from vertex functions $p_i : [-1,1]\rightarrow \mathbb{R}^d$.
Since the $p_i$ are generally \textit{inconsistent} across overlaps, we enforce consistency by blending adjacent vertex functions.
We define non-overlapping `edge-based local maps' by blending between the vertex functions at each end of the edge:
\begin{equation}
         c_{i,i+1}(t) := p_i(t)\,B(t) 
         + p_{i+1}(\tau_{i,i+1}(t))\,B(\tau_{i,i+1}(t)) \; \text{for} \; t \in (0,1].
     \label{eq:edge-c}
\end{equation}
Then, we can define overlapping vertex-based local maps $c_i$ by,
\begin{equation}
     c_i(t) = \left \{
     \begin{matrix}
        c_{i,i+1}(t)\; \text{for} \; t \in (0,1] \\[1mm]
         c_{i-1,i}(   \tau_{i,i-1}  ( t) ) \; \text{for} \; t \in [-1,0],
     \end{matrix} 
     \right.
     \label{eq:vtx-based-curve}
\end{equation}
%
%
%
where $B:[-1,1] \rightarrow [0,1]$ is a \textit{blending function} (see \Cref{fig:1d_blending}).
These $c_i$ satisfy the  defined by equation \ref{equation:consistency} by construction: for $t \in [0,1]$ then
$c_{i+1}(\tau_{i,i+1}(t)) = c_i(t)$, and similarly for $t \in [-1,0]$,
$c_{i-1}(\tau_{i,i-1}(t)) = c_i(t)$.

We refer to a collection of such local maps as a \textit{blended chart curve}. It has a \emph{local control} property: modifying the vertex function $p_i$ affects only $c_i$, $c_{i-1}$, and $c_{i+1}$.
Additionally, the blended chart curve should ideally be (i) at least $C^\infty$ continuous everywhere; (ii) equivariant to rigid transforms and scaling of the coarse curve.

\begin{figure*}[t!]
    \centering
    \includegraphics[width=\linewidth]{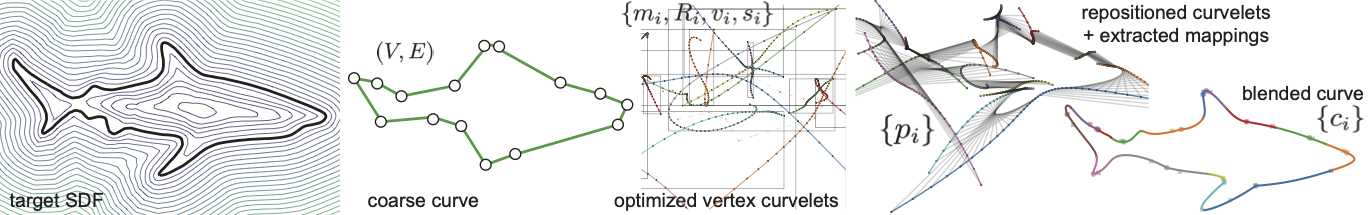}
    \caption{\textbf{Blended chart curve fitting}. (Left to right) We start from the target implicit field (closed-curve SDF) and a coarse proxy polygon $(V,E)$. At each coarse vertex, we optimize a local curvelet parametrized by maps $m_i$ (in this example, degree-5 polynomials) together with their associated local frame (i.e, rotation $R_i$, translation $v_i$, scale $s_i$). These optimized curvelets are then repositioned to form the vertex functions $\{p_i\}$, which are finally combined via the blending functions to produce the blended chart curve $\{c_i\}$ defined over the coarse proxy curve. The blended chart curve is smooth by construction and, more importantly,  equivariant to rigid motions and scaling of the coarse proxy.}
    \label{fig:curve_shark}
\end{figure*}

\paragraph*{Smoothness conditions.}
We would like for the curve to possess $C^\infty$ continuity, in the sense that the local parametrizations (from the domain intervals) are $C^\infty$, and the transition maps are $C^\infty$ (similarly to how a $C^\infty$ manifold is defined in Chapter 1, p14-15 of \cite{lee2013}).

We require that each $p_i$ is $C^\infty$, and that the blending function $B$ satisfies:
\begin{enumerate}[label=(C\arabic*), leftmargin=*]
    \item $B$ is $C^\infty$ continuous.
    \item $B(\pm 1) = 0$.
    \item $\dfrac{d^rB}{dx^r}(\pm 1) = 0$ for all $r\geq 1$ 
\end{enumerate}
Property (C1) ensures that each $c_i$ is $C^\infty$ on $(-1,0)$ and $(0,1)$ (via the Product Rule).
Property (C2) ensures that the left and right limits of $c_i$ agree at $t=0$. Property (C3) ensures agreement of the relevant higher derivatives at $t=0$ (see the supplemental for details).

Note that a linear blending function (e.g., the Lagrange hat function) is not suitable because it has only $C^0$ continuity at zero and $\pm 1$, which means that the $c_i$ functions will have sharp bends at $t=0$ (for almost all $p_i$ functions).

\paragraph*{Equivariance via proxy geometry dependence.}
To ensure equivariance of the vertex functions $p_i$ to translation, rotation ($R_i$), and scaling ($s_i$) of the coarse control mesh, we construct $p_i$ in terms of the associated coarse vertex position $v_i$, as: 
\begin{equation}
p_i(t) = s_i R_i m_i(t) + v_i,
\label{eq:coarse_dep}
\end{equation}
where $m_i: \mathbb{R}\rightarrow \mathbb{R}^2$ controls the specific shape of the function. (In Figure \ref{fig:curve_shark}, degree-5 polynomials are used.)
We set the $2\times 2$ rotation matrix, 
\begin{equation}
R_i =
\begin{pmatrix}
\mathbf{N}_i & \mathbf{N}^\perp_i
\end{pmatrix},
\end{equation}
where $\mathbf{N}_i$ is the $i$\textsuperscript{th} coarse vertex normal, and
\begin{equation}
    s_i = s_\text{global}\cdot \frac{1}{2}
    \left( \left \| v_{i+1} - v_i \right \|
         + \left \| v_i - v_{i-1} \right \| \right).
\end{equation}
The parameter $s_\text{global}$ is fixed during initialization. 
The $p_i$ vertex functions are equivariant to scaling and rotation of the proxy geometry, and this carries forward to the blended maps $c_i$ (see supplemental B.1).
Finally,  $c_i$ inherits translation equivariance from the $p_i$ if and only if $B$ satisfies \textit{partition of unity} (see supplemental B.2):
\begin{enumerate}[label=(C\arabic*), leftmargin=*]
\setcounter{enumi}{3}
    \item $B(t) + B(\tau_{i,i-1}(t)) = 1 \quad \text{for} \quad t \in [-1,0],$\\
          $B(t) + B(\tau_{i,i+1}(t)) = 1 \quad \text{for} \quad t \in (0,1].$
\end{enumerate} 

\paragraph*{Choice of blending function.}
We build a blending function satisfying (C1)--(C4) from an exponential function.
Let
\begin{equation}
f(t) =
\left\{
\begin{matrix}
 \exp\!\left(-\dfrac{1}{t}\right) \quad \text{for} \quad t>0,\\
 0 \quad \text{otherwise.}
\end{matrix}\right.
\label{eq:f-defn}
\end{equation}
This is infinitely differentiable  (see \cite{Nestruev2003SmoothManifolds}, Chapter 2).
Then, 
\begin{equation}
g(t) = \frac{f(1-t)}{f(t)+f(1-t)}
\label{eq:g-defn}
\end{equation}
is a smooth transition function between one and zero.
Finally, we define the blending function to be,\\
\begin{equation}
    B(t) := g \left( \frac{|t|- (1-\beta)/2  }{\beta} \right).
    \label{eq:B-defn}
\end{equation}
The parameter $\beta \in (0,1)$ controls the length of  the overlap region 
\begin{wrapfigure}[8]{r}{0.5\columnwidth}
\vspace*{-.1in}
  \begin{center}
\hspace{-.3in}
    \includegraphics[width=0.45\columnwidth]{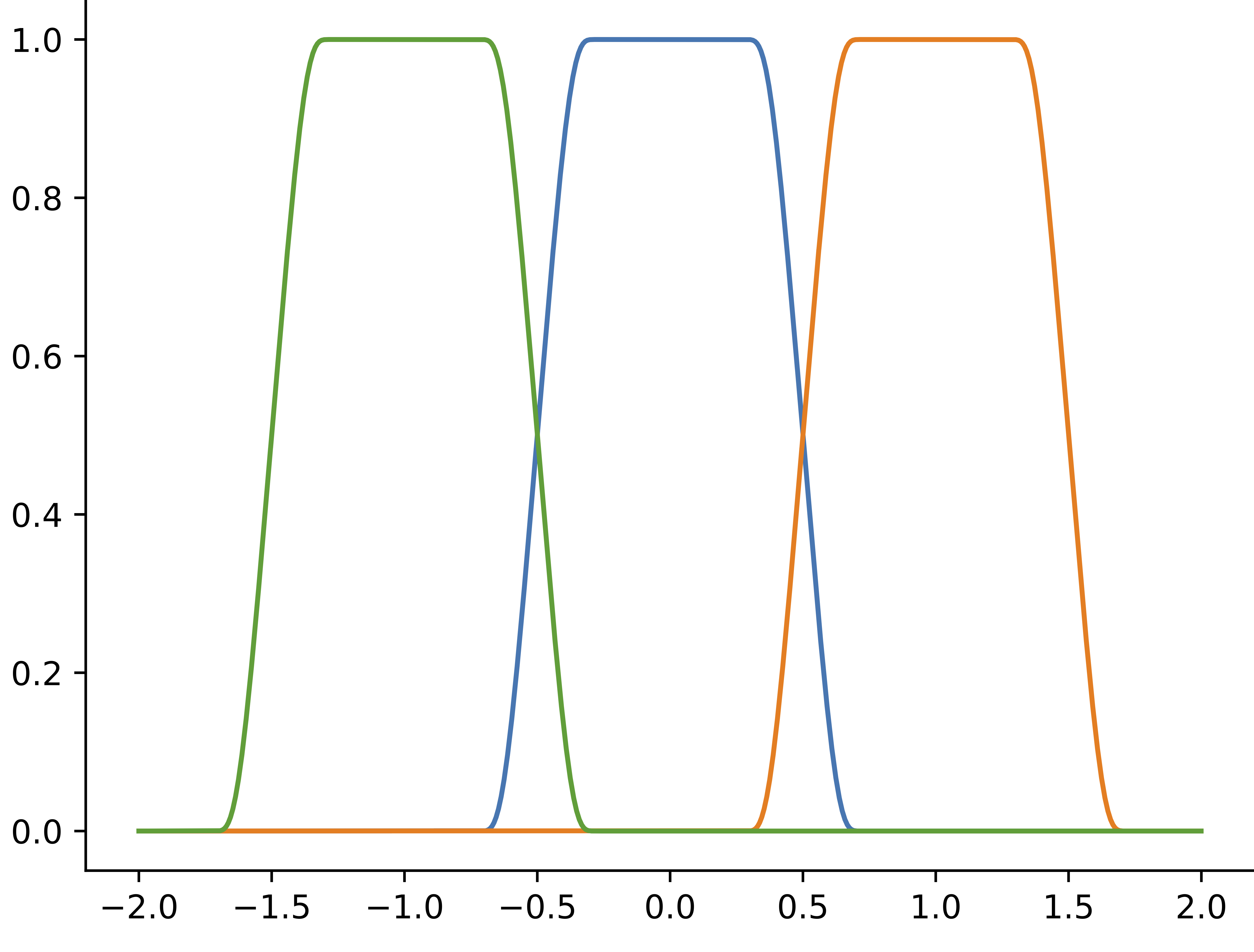}
  \end{center}
\end{wrapfigure}
 where both components are positive (see inset showing blending functions $B(t)$ with $\beta=0.5$). We demonstrate the $C^\infty$ continuity of $B$, and of the Blended Chart Curve, in Supplemental sections B.3 and B.4.

\paragraph*{Optimizing the vertex functions.}
We define the maps $m_i$ parametrically (e.g., polynomials with optimizable coefficients), and optimize these coefficients directly, using Adam~\cite{adam}, to make a blended chart curve approximate a curve given in the form of an implicit function.

In order to convert a level-set (implicit) representation of a curve, such as a signed distance field~(SDF), into a blended chart curve (an explicit representation), we manipulate the implicit representation into a differentiable objective whose minimum occurs at the zero level set.
For example, we minimize the absolute SDF (i.e., the UDF or `Unsigned Distance Function') values, 
\begin{equation}
   \mathcal{L}_\text{SDF} :=  \frac{1}{n} \sum_{i=0}^{n-1}\int_{0}^{1} |\textbf{sdf}(c_{i,i+1}(t))| dt.
\end{equation}
Minimizing this objective ensures that the image of the blended chart curve lies on the intended curve, but does not ensure that it covers the whole curve.

In \Cref{fig:curve_shark}, we use degree-5 polynomials as the $m_i$ functions. 
In practice, we found that to prevent collapse to non-surjective solutions, it is sufficient to set $s_\text{global}$ to a small value (0.1). This biases the optimization away from producing $m_i$ functions with large outputs, thereby softly preventing the blended chart curve from straying too far from the coarse curve.

\paragraph*{Differential quantities and regularization.}
We compute exact differential quantities at any point on the blended chart curve using \texttt{autograd}. 
We can further regularize the blended chart curve using first-order properties. Close to the true curve, the SDF gradient aligns with the outward normal. To discourage foldovers, we penalize misalignment between the blended chart curve normal and the SDF gradient, 
\begin{equation}
    \mathcal{L}_\text{normal} := \frac{1}{n}\sum_{i=0}^{n-1} \int_0^1 \left[1 - \nabla \textbf{sdf}(c_{i,i+1}(t)) \cdot \mathbf{n}_{i,i+1}(t)\right]\,dt.
\end{equation}
This loss is minimized when the blended chart curve normals align exactly with the SDF gradient. If the Eikonal property is satisfied (i.e., $\left \| \nabla \textbf{sdf} \right \| = 1 $ everywhere) then this minimum is zero. Note that if we use a level-set representation that is not a true SDF, then the gradient does not have unit-norm everywhere, so the loss value at this minimum could be above or below zero.

\begin{figure}[b!]
    \centering
    \includegraphics[width=\columnwidth]{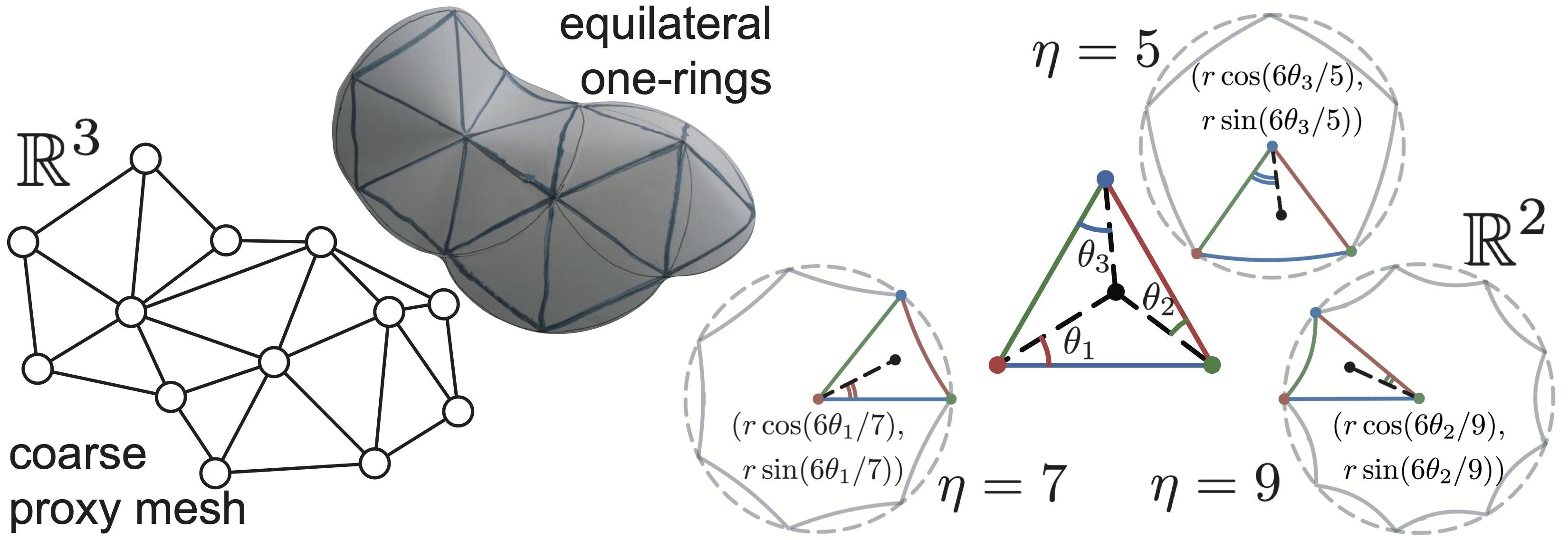}
    \caption{\textbf{Computing one-ring coordinates.} Our one-ring coordinates are based on the idea that if we affinely transform every triangle in a small part of a mesh into a unit equilateral triangle, without changing the connectivity, then the one-rings in this transformed mesh-part would all look like (be isometric to) cones---including hyperbolic cones, where $\eta > 6$, planar parts, where $\eta=6$, elliptic points, where $\eta < 6$. The position of a point on a cone can be described by the distance $r$ to the tip, and the angle $\theta$ around the cone from a fixed edge.   Every triangle belongs to three one-ring `cones'. These cones can be flattened to unit discs, by multiplying the angle $\theta$ by $6/\eta$. The `one-ring-coordinates' of a point are coordinates of where the point ends up in the three different discs.
    We illustrate a point on the abstract equilateral triangle mesh-part, its angles $\theta_i$ (modulo $\pi/3$), and the representation of this point in the one-ring-coordinates of the three vertices (valences $5$, $7$, and $9$).
    }
    \label{fig:one-rings}
\end{figure}

\subsection{Blended Chart Surfaces (2D)}

Our goal is to build a smooth surface, a \textit{Blended Chart Surface}, over a coarse triangle mesh. We proceed analogously to the curve case.

\paragraph*{Local maps and one-ring coordinates.}
We associate a $C^\infty$-continuous local map $p_i$ with each vertex $v_i$. To ensure equivariance to rotation and scaling of the proxy, $p_i$ depends on $m_i$ via the 3D version of Equation \ref{eq:coarse_dep}.
Recall that in the curve case, the domain of $p_i$ was $[-1,1]$, which can be viewed as two unit-length edges joined at $0$ and ``flattened'' into 1D.
Analogously, in the surface case, the domain consists of all triangles in the one-ring of vertex $i$, \textit{affinely} deformed to unit-side-length equilateral triangles and then ``flattened'' into 2D.
Every point in this \emph{equilateral one-ring} (before flattening) admits a polar representation $(r,\theta)$, where $r$ is the distance to the central vertex and $\theta \in [0,\eta\pi/3)$ is the oriented angle from a fixed reference edge for a vertex of valence $\eta$ (see \Cref{fig:one-rings}).
The flattening to $\mathbb{R}^2$ preserves the radial distance and rescales angles so that the range becomes $[0,2\pi)$. Thus, we have, 
\[
(r,\theta) \mapsto \bigl(r\cos(6\theta/\eta),\; r\sin(6\theta/\eta)\bigr).
\]

We refer to these flattened coordinates as \emph{one-ring coordinates}; these are the coordinates used as input to the vertex functions $m_i$. Our flattening map is similar to the `conformal flattening map' from the splines literature (see for example \cite{interpolatingSplines}). These flattening maps are equivalent in the case where a one-ring consists of unit-length equilateral triangles.

To be precise, we take a one-ring to \textit{not} include its outer boundary (this is also known as the ``star'' of the vertex, see~\cite{munkresAlgTop}). Thus, every point on the mesh belongs to three one-rings (if it lies on the interior of a triangle), two one-rings (if it lies on the interior of an edge), or one one-ring (if it is a vertex). See \Cref{fig:one-rings} for an illustration of the three sets of one-ring coordinates for a point inside a triangle of a coarse mesh.

\begin{figure}[b!]
    \centering
\includegraphics[width=\columnwidth]{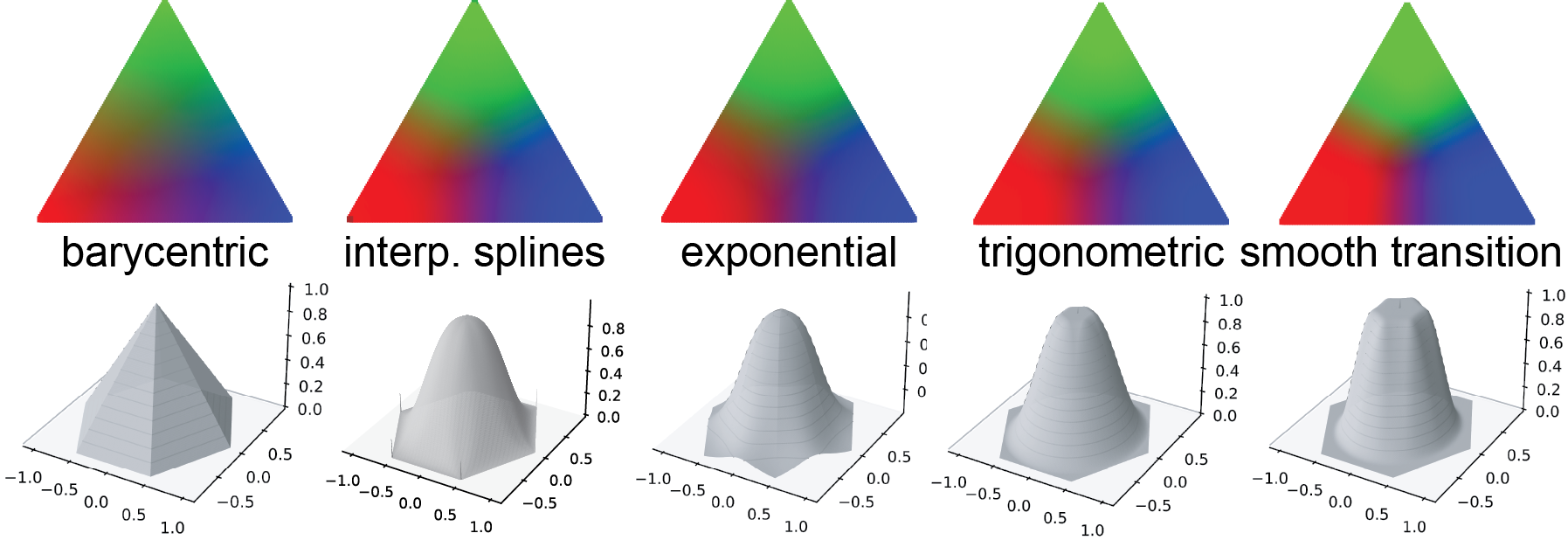}
    \caption{\textbf{Choices of blending functions} on an equilateral triangle and their induced vertex-centered `hat' functions. (Top) Weight distributions over the equilateral triangle: barycentric interpolation and three other variants based on radial distance (exponential, trigonometric, and smooth transition). See Section 4 for the definitions.
    (Bottom) The corresponding blending function around a valence-$6$ vertex, illustrating how the choice of blending function affects the smoothness and support of blending around the vertex.}
    \label{fig:blending-1d-2d}
\end{figure}

\begin{figure*}[t!]
   \includegraphics[width=\linewidth]{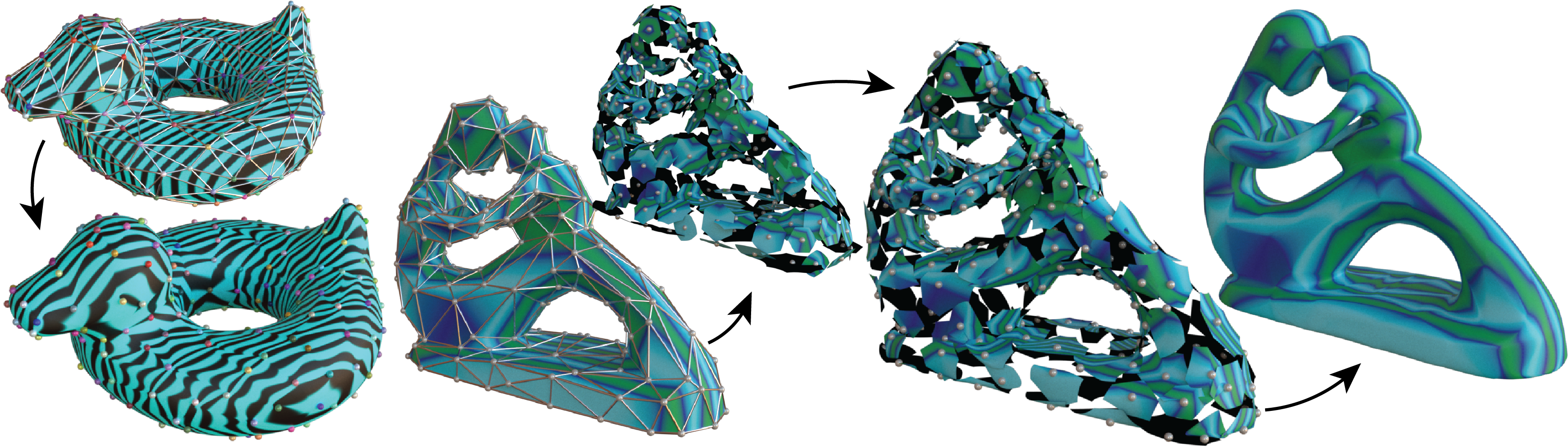}
\caption{\textbf{Discovering a surface parametrization.} \name implicitly induce a seamless correspondence between the proxy and the fitted surface via the one-ring coordinates and blended local maps. {Left (Bob):} We visualize the optimized correspondence by pulling back a procedural color pattern defined on the coarse proxy mesh onto the Blended Chart Surface. {Right (Fertility):} We illustrate how a color field is transferred through the different stages: an input coloring on the proxy is mapped to the unblended polynomial patches (each point on a coarse triangle contributes to three local patches), then pushed through the optimized patches, and finally transferred to the BCS.}
    \label{fig:parameterization}
\end{figure*}

\paragraph*{Transition maps.}
The transition maps convert between pairs of one-ring coordinate systems, where two one-rings overlap. These transition maps are diffeomorphisms (see the supplemental) 
which is crucial for ensuring that the Blended Chart Surface has guaranteed smoothness.

\paragraph*{Blending vertex functions.}
As in the curve case, we turn an arbitrary set of local maps $p_i$ into a consistent set by blending them together.
If $\mathbf{x}$ is a point in the equilateral triangle $ijk$, we define the face-based local map as, 
\begin{equation}
     c_{ ijk} (\mathbf{x}) =
     \begin{matrix}
         p_i(\texttt{oc}^{ijk}_i(\mathbf{x}))B^{\text{2D}}(\mathbf{x}, i)  
         +   p_j(\texttt{oc}^{ijk}_j(\mathbf{x}))B^{\text{2D}}(\mathbf{x}, j) 
          \\ + p_k(\texttt{oc}^{ijk}_k(\mathbf{x}))B^{\text{2D}}(
         \mathbf{x}, k)
     \end{matrix} 
     \label{eq:consistent_triangle_map}
\end{equation}
where $\texttt{oc}^{ijk}_i(\mathbf{x})$ denotes the one-ring coordinates of $\mathbf{x}$ (in triangle $ijk$) in the one-ring centered at vertex $v_i$; the function $B^{\text{2D}}(\cdot,i)$ is a blending function defined on the canonical equilateral triangle domain $T$, which equals $1$ at vertex $i$ and $0$ at the other two vertices. Notice that the vertex functions are applied to one-ring coordinates, while the blending functions are applied directly to the equilateral triangle coordinates.

Analogously to \Cref{eq:vtx-based-curve} where edge-based local maps were joined together to create consistent overlapping vertex-based local maps, we can imagine grouping together all of the $c_{ijk}$ maps for triangles $ijk$ in a one-ring, to produce vertex-based maps, $c_i : \text{one-ring}_i \rightarrow \mathbb{R}^3$, that are consistent on their overlaps.

\paragraph*{Conditions on the 2D blending function.}

We would like for the surface to possess $C^\infty$ continuity, in the sense that the local parametrizations from one-ring coordinates are $C^\infty$, and the transition maps are $C^\infty$ (similarly to how a $C^k$ manifold is defined in Chapter 1, p14-15 of \cite{lee2013}).

We require $B^{\text{2D}}$ to satisfy:
\begin{enumerate}[label=(S\arabic*), leftmargin=*]
  \item $B^{\text{2D}}$ is $C^\infty$ continuous.
  \item $B^{\text{2D}}(\mathbf{x},i)=0$ in a neighborhood of the edge opposite to vertex $i$.
  \item The directional derivatives perpendicular to the edges are zero along all edges (and the derivatives at the vertices are all zero).
  \item $B^{\text{2D}}(\mathbf{x}, i) + B^{\text{2D}}(\mathbf{x}, j) + B^{\text{2D}}(\mathbf{x}, k)=1$ (partition of unity).
\end{enumerate}
Property (S1) ensures that each $c_{ijk}$ is $C^\infty$ on the interior of the triangle, provided the $p_i$ are sufficiently smooth.
Properties (S2) and (S3) ensure that when we form hat functions on the mesh by combining the per-triangle blending functions within a one-ring, and assigning value zero outside the one-ring, the resulting hat function has $C^\infty$ continuity across edges between triangles.

Finally, we construct the 2D blending function $B^{\text{2D}}$ in terms of a 1D blending function $B$ via 
(see \Cref{fig:blending-1d-2d}), 
\begin{equation}
B^{\text{2D}}(\mathbf{x}, i) =
\frac{B(|\mathbf{x} - v_i|)}
     {B(|\mathbf{x}-v_i|) + B(|\mathbf{x}-v_j|) + B(|\mathbf{x}-v_k|)}.
\label{eq:2d_blend_func}
\end{equation}
As described in \Cref{subsec:BPC}, the 1D blending function we chose has an overlap parameter $\beta$, and it reaches zero at $\alpha = (\beta+1)/2$.
For (S2) to be satisfied, we require $\alpha < \sqrt{3}/2$ (see Supplemental B.5).
For $B^{\text{2D}}$ to be well-defined, the denominator of \Cref{eq:2d_blend_func} must be non-zero for all $\mathbf{r}$, which yields the condition $\alpha > \sqrt{3}/3$ (see Supplemental B.5).
Thus, $B$ reaches zero before the one-ring boundary, and there is no empty region where all blending functions vanish (see supplemental for details).
We choose $\beta=0.73$, hence $\alpha=0.865$, which is slightly smaller than $\sqrt{3}/2$.


\paragraph*{Coarse-vertex dependence and equivariance.}
Analogously to the curve case, we make each $p_i$ depend on the coarse vertex positions via \Cref{eq:coarse_dep}.
For surfaces in 3D, $R_i$ is a local coordinate frame (a $3 \times 3$ rotation matrix) aligned to the $i^{\text{th}}$ vertex normal and the tangent plane direction closest to an arbitrary edge of the one-ring, and $s_i$ is a local scale (proportional to the mean length of edges emanating from vertex $i$).
Rotation and scale equivariance follow from \Cref{eq:coarse_dep}, and translation equivariance follows from partition of unity, which is enforced by construction in \Cref{eq:2d_blend_func}.

\begin{figure*}[t!]
\includegraphics[width=\linewidth]{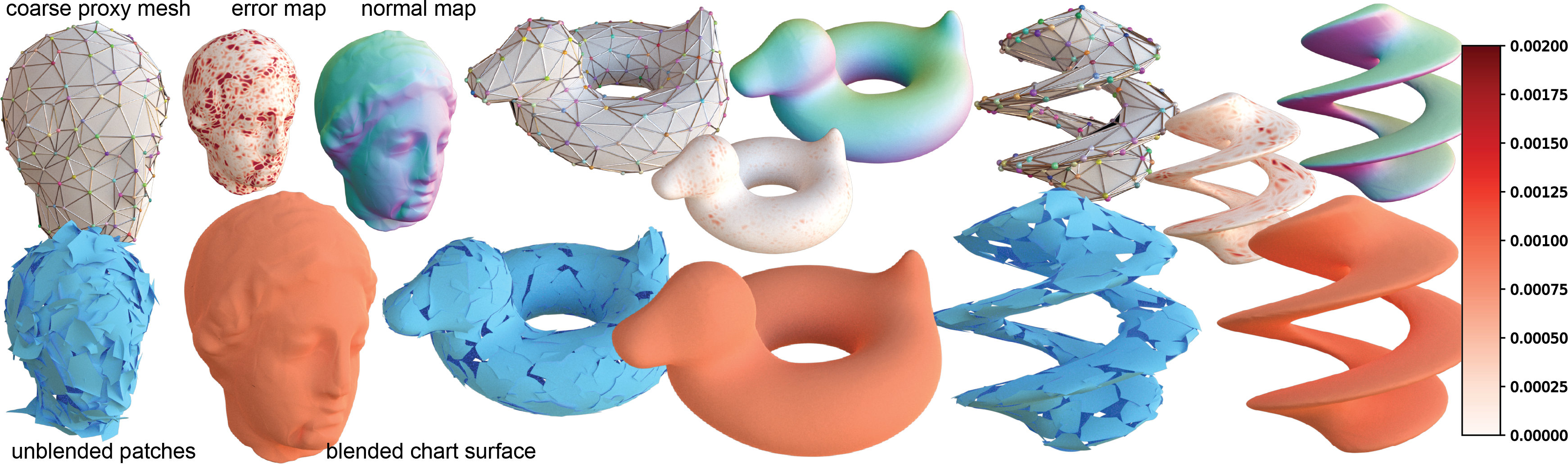}
\caption{\textbf{Result gallery.} BCS on different targets (Igea, Bob, and a twisted torus). For each model, we show the coarse  proxy mesh, the optimized but \emph{unblended} local polynomial patches (illustrating the patch structure prior to blending; quadratic polynomials used in these examples), the resulting Blended Chart Surface, and the corresponding error map (color bar at right; models are normalized to a unit-width bounding box). We also visualize the normal map computed using Blended Chart Surface to highlight stable, smoothly varying normals. All results use quadratic patches (i.e., $18$ scalar coefficients per vertex), and the proxy meshes have 252, 250, and 200 vertices, respectively.}
    \label{fig:results}
\end{figure*}

\paragraph*{Fitting to implicit targets.}
Instead of pre-computing a local or global parametrization of the target shape, we use the same technique described in Section~\ref{subsec:BPC}: we take an \textit{implicit} representation of the target shape and optimize the parameters of the $m_i$ functions, using Adam~\cite{adam}, to fit the zero level set.

We choose the $m_i$ functions to be quadratic polynomials in the one-ring coordinates:
\[
m_i(u,v) = A_i \, \Phi(u,v)^T \text{where}
\]
\[
\Phi(u,v) =
\begin{bmatrix}
1 & u & v & u^2 & uv & v^2
\end{bmatrix},
\]
where $A_i$ is a $3 \times 6$ matrix (for a quadratic polynomial), for vertex $i$ to be optimized.

If, for example, the implicit function is an SDF, we minimize the mean absolute value of the SDF over the surface:
\begin{equation}
\mathcal{L}_\text{SDF} :=
\frac{1}{|F|} \sum_{ijk \in F } \int_T  w_{ijk}(\mathbf{x})
  |\textbf{sdf}(c_{ijk}(\mathbf{x}))| d\mathbf{x}.
\end{equation}
However the implicit function does not need be an SDF; we also show examples on deformed SDFs (e.g. Figure \ref{fig:remeshing}), neural implicits (e.g. on Igea, Bob and Fertility), and UDFs (e.g. on the M\"obius band).

The domain $T$ is the canonical equilateral triangle with vertices $(0,0)$, $(1,0)$, and $(0.5,\sqrt{3}/2)$.
For a fast approximation, the weight function $w$ can be taken to be one everywhere; for a more accurate result it can be taken proportional to local area distortion, computed via the determinant of the first fundamental form. See the supplemental for how to precompute derivatives for efficient computation. 

\section{Evaluation}
\label{sec:evaluation}

\paragraph*{Experimental setup.}
\update{Our method supports various forms of implicit fields and distance functions, as long as they are differentiable. We evaluate \name on two classes of targets: (i)~learned implicit functions (neural SDFs), and (ii)~analytical implicit functions (not necessarily true SDFs).

(i) \textbf{Neural SDF}: For the results on ‘Fertility’, ‘Bob’ and ‘Igea’ the input field is an overfitted neural network. The network consists of a learnable latent, a transformer encoder, and a cross-attention mechanism to evaluate the function at Fourier-Embedded 3D coordinates and GeLU activations. 
(Note, for the 2D shark example in Figure 4, the input is a simpler neural SDF, represented by a 4-layer MLP with 128 nodes in each layer and ReLU activations.)

(ii) \textbf{Analytical implicits}: The input for the torus (Figure \ref{fig:elastic}) is an analytic SDF. The input for the shapes in Figures \ref{fig:motivation} and \ref{fig:remeshing} are analytic implicit fields (not strictly true distance fields), designed by taking the analytic SDF of a sphere and torus, respectively, then adding a sinusoidal term. 
The input for the Möbius band (Figure \ref{fig:mobius}) is an analytical Unsigned Distance Function (UDF), because the surface is non-orientable.
}

\paragraph*{Proxy meshes.}
For signed implicit fields, we obtain proxy meshes by applying quadric edge-collapse decimation with topology preservation to a Marching Cubes extraction. For the M\"obius band, which is specified only by an unsigned implicit field, we construct the proxy mesh analytically. The Igea, Fertility, Bob, and Twisted Torus examples use 500 faces. The M\"obius band proxy has 220 faces (165 vertices). \Cref{fig:motivation} uses 210 faces;  \Cref{fig:elastic} uses 100 faces.


\paragraph*{Optimization and losses.}

The main examples are optimized in the fast setting, i.e., without area-weighted sampling, normal regularization, or distortion regularization. 
By default, we use quadratic polynomials ($6 \times 3$ coefficients per vertex). The blending-function is the `smooth transition function' with overlap parameter is $\beta=0.73$ unless otherwise specified, and we set $s_\text{global}=0.5$, except for the Igea example in Figure~\ref{fig:results}, which uses $s_\text{global}=0.7$ (visible in the size of the blue, unblended patches). We run the Adam optimizer directly on the polynomial coefficients, for up to 20 epochs.
\update{The learning rate starts at 1e-3, decreasing by cosine annealing to 1e-5 over 20 epochs. We sample and cache  10,000 samples uniformly on each domain triangle, and in each batch we process a random selection of 100 samples per triangle.}

Runtime is approximately 10--15 minutes in the fast setting (positional loss only). It takes about 10 minutes with the trigonometric blending function, and about 15 minutes with our proposed smooth-transition blending function. The suggested regularizers (normal loss, distortion loss, and area-weighted sampling) take longer per batch as the loss function then involves deriatives. With the proposed blending function and normal regularization, runtime is approximately an hour to converge. All experiments are run on a MacBook Pro M3 laptop with 96GB RAM.

\begin{figure}[t!]
    \centering
    \includegraphics[width=\columnwidth]{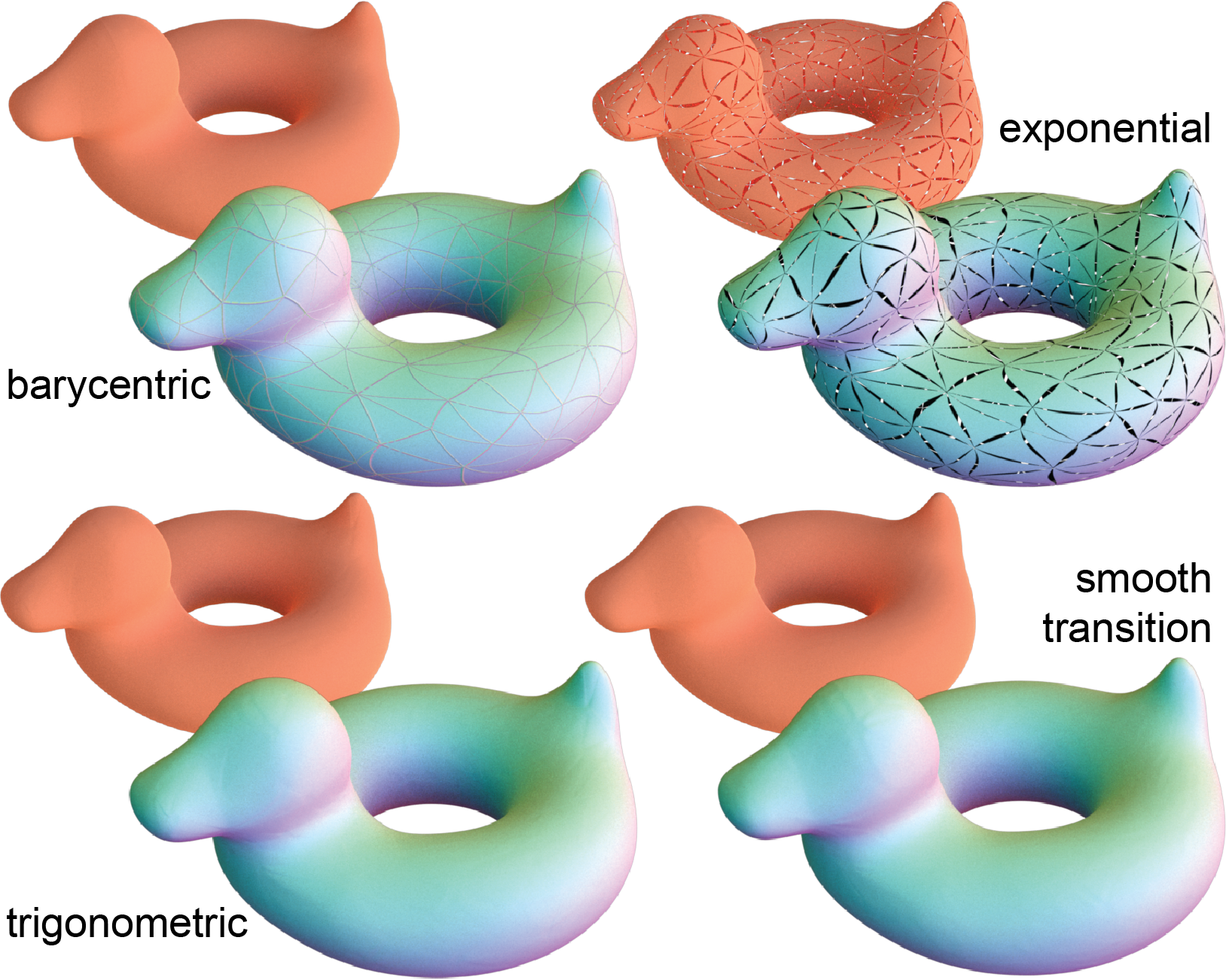}
    \caption{\textbf{Effect of the blending function} on continuity and surface quality (see \Cref{fig:blending-1d-2d}). {Barycentric} blending yields only $C^0$ continuity, as normals are not well-defined along coarse mesh edges. A direct {exponential} weighting fails to preserve even $C^0$ continuity at edges, producing visible holes. {Trigonometric} blending achieves $C^1$ continuity and yields visually smooth surfaces. Our {smooth-transition} blending is $C^\infty$ and produces globally smooth geometry.
    }
    \label{fig:blendingContinuity}
\end{figure}

\paragraph*{Baselines.}
We compare three alternative blending functions to our proposed smooth-transition blending: barycentric, exponential, and trigonometric (\Cref{fig:blendingContinuity} and \Cref{fig:blending-1d-2d}). The blending functions are: \\
(i) \textit{Barycentric interpolation:} 
\begin{equation}
        B^{2D}(\mathbf{x}, i) = \frac{\text{Area}(\mathbf{x}, v_j, v_k)}
            {\text{Area}(\mathbf{x}, v_j,v_k) + 
             \text{Area}(\mathbf{x}, v_i,v_j) +
             \text{Area}(\mathbf{x}, v_k, v_i)}
\end{equation} \\ 
(ii) \textit{Exponential interpolation:} 
\begin{equation} B^{2D}(\mathbf{x}, i) = \frac{ h^{|\mathbf{x} - v_i|} }
 {  h^{|\mathbf{x} - v_i|} + h^{|\mathbf{x} - v_j|} + h^{|\mathbf{x} - v_k|}     }   
\end{equation} \\ 
(The value $h$ must be a small value in $(0,1)$, and we chose $h=0.01$.) \\ 
(iii) \textit{Trigonometric interpolation:} see  \Cref{eq:2d_blend_func},  with the radial function  $B$ defined by 
\begin{equation}
        B(r) =  \text{trig} \left( \frac{r-a} {1 -2a } \right)
\end{equation}
where
\begin{equation}
        \text{trig}(t) = \text{cos} \left ( \frac{\pi}{2} \text{clamp}_{[0,1]}(t) \right ) ^2 
\end{equation} and $a = 0.5 - \beta/2$. \\
(iv) \textit{Smooth transition interpolation:} 
our proposed blending functions are in Equations 
\ref{eq:f-defn}, \ref{eq:g-defn}, \ref{eq:B-defn} and 
\ref{eq:2d_blend_func}.

Barycentric blending violates Properties S2 and S3, resulting in gradient discontinuities aligned with proxy-mesh edges, which appear as sharp creases. In other words, the \textit{barycentric} choice leads to a `hat' with tangent discontinuities,  propagating to the blended chart surface. The exponential blending function strongly violates Property S2: it does not reach zero at the boundary of the one-ring, leading to clear discontinuities (holes) in the Blended Chart Surface. Trigonometric blending function produces similar visual quality to our method, but yields only $C^1$ continuity rather than $C^\infty$. It is also slightly faster to optimize, so trigonometric blending can be a practical alternative within our framework when only visual smoothness is required. See \Cref{fig:blendingContinuity}.

The \textit{interpolating splines} blending function (used by \protect{\cite{interpolatingSplines}}) has $C^2$ continuity on the equilateral one-ring shown, but does not strictly fit into our framework because the blending function is not affine equivariant, since it depends on both barycentric and 3D coordinates; empirically, it produces tangent discontinuities when applied to vertex functions that use our one-ring-coordinates.

As another baseline, we consider a simple MLP displacement field (in $\mathbb{R}^3$) defined on a coarse proxy mesh (two hidden layers with 64 units and SiLU activations). While the MLP defines a smooth field, the proxy geometry contains sharp edges, which remain visible after deformation. Note that this representation is not equivariant to the base mesh. 
This experiment emulates methods such as NGFM~\cite{NeuralGeomFields}, which also deform coarse proxy geometry with a neural network. NGFM uses bilinear quadrilateral faces rather than triangles and lifts points to a higher-dimensional feature space; but sharp edges can still appear where bilinear patches meet, so the underlying discontinuity issue remains.


\paragraph*{Qualitative results.}
We show representative shapes spanning different topologies and geometric complexity, for both analytical and neural implicit targets, in Figures \ref{fig:teaser} and \ref{fig:results}. Unlike methods relying on global parametrization~\cite{morreale2021neural,williamson2025spherical}, our approach readily handles higher-genus surfaces (Bob and Twisted Torus are genus 1; Fertility is genus 4). Error maps visualize $|\textbf{sdf}|$ at each surface point (white = low, red = high; colormap capped at 0.002 after unit-box normalization). Igea, Bob, and Fertility are fitted to neural SDFs. Twisted Torus is fitted to an analytic implicit field (a deformed torus SDF that is no longer strictly an SDF), which we also use for the error map. Overall, reconstructions are strongest on smoother shapes (Bob, Twisted Torus) and slightly less accurate on highly detailed geometry (Igea); fidelity can be improved by refining the proxy or increasing the expressivity of the vertex functions.

\begin{figure}[t!]
   \includegraphics[width=\columnwidth]{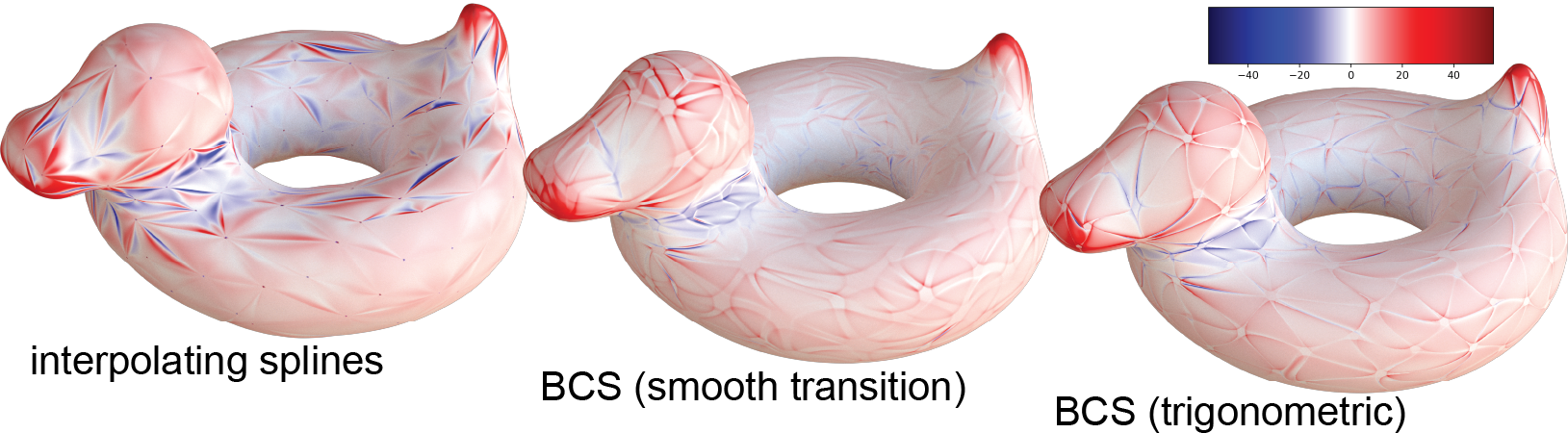}
    \caption{\textbf{Mean Curvature} colormaps computed using automatic differentiation on a BCS with `smooth transition' blending and a BCS with trigonometric blending, and discrete mean curvature computed on a high-resolution mesh of an Interpolating Spline~\cite{interpolatingSplines}, using the same coarse net and the same class of vertex functions (quadratic polynomials).    }
    \label{fig:curvature}
\end{figure}

\begin{figure}[b!]
    \includegraphics[width=\columnwidth]{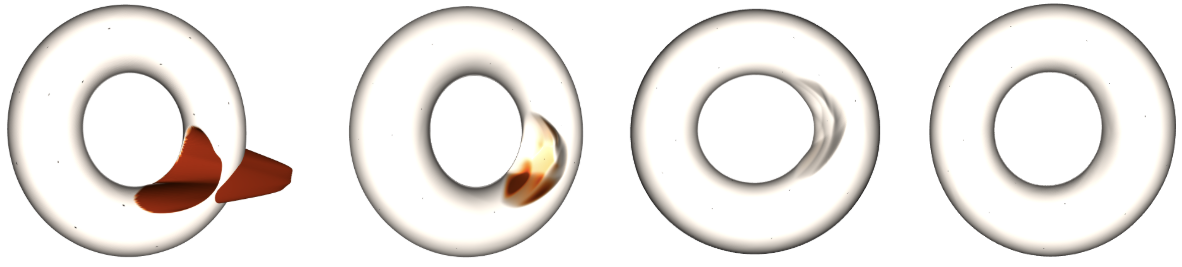} 
    \caption{\textbf{Elastic energy.} We displace one proxy vertex of a \name torus while keeping the vertex functions fixed. We compute the deformation gradient from the Jacobians of the deformed and undeformed tori, and evaluate elastic energy via the Green--Lagrange strain (rubber-like material parameters). Red indicates high energy density and white indicates zero. We then reoptimize the vertex functions with Adam to minimize total elastic energy, and the surface relaxes back toward its original shape. This example highlights the potential of \name for differentiable physical simulation, which is difficult to achieve directly with implicit/volumetric representations.  }
    \label{fig:elastic}
\end{figure}

\paragraph*{Access to surface properties.}
Blended Chart Surfaces support stable evaluation of differential quantities. We compute exact normals continuously as the cross product of automatic derivatives with respect to the two coordinates of the canonical equilateral triangle domain; these normals are shown in Figures \ref{fig:teaser}, \ref{fig:results}, \ref{fig:blendingContinuity}, \ref{fig:remeshing} and \ref{fig:truncation} (left).
We can also compute curvatures on the surface; Figure~\ref{fig:curvature} shows mean curvature values.

Figure \ref{fig:elastic} illustrates computing and optimizing elastic energy, as another differential quantity, highlighting the potential for downstream  physical simulation.
We displace one proxy vertex of a Blended Chart Surface torus while keeping the vertex functions fixed. We compute the deformation gradient from the Jacobians of the deformed and undeformed tori, and evaluate elastic energy via the Green--Lagrange strain (rubber-like material parameters). We  re-optimize the vertex functions with Adam to minimize total elastic energy, and the surface relaxes back towards its original shape. This example highlights the potential of Blended Chart Surfaces for differentiable physical simulation of thin sheets, which is difficult to achieve directly with implicit/volumetric representations. 
Modeling thin sheets is natural in our framework because the representation is inherently 2D, not volumetric. We have access to local parametrizations (the equilateral one-rings); pulling these back to the coarse mesh yields a correspondence between the coarse and blended geometries (\Cref{fig:parameterization}), though it is not globally smooth since each coarse triangle is affinely transformed to an equilateral one.

Figure \ref{fig:curvature} shows mean curvature computed on Blended Chart Surfaces~(BCS) with our `smooth transition' blending function and `trigonometric' blending, via the Second Fundamental Form (using automatic differentiation), and mean curvature on an Interpolating Spline~\cite{interpolatingSplines} on the same proxy mesh (found using a standard discrete method: applying the cotan Laplace Beltrami operator to the matrix of vertex-positions). The curvature map exhibits distinctive behaviour: the `trigonometric' BCS has discontinuous curvature showing that the surface is $C^2$ not $C^3$ or higher, the `smooth transition' BCS has continuous curvature but shows structured `stripes' of flatter curvature near edges, and the Interpolating Spline has poor curvature quality because the surface is not fitted, but it does not suffer from the `structured' artifacts present in ours, which we attribute mainly to the differences in the shape of the blending `hat' function (Figure \ref{fig:blending-1d-2d}).

\paragraph*{Effect of Proxy Mesh.}
Blended Chart Surfaces inherit the proxy topology, so the proxy must have the correct topology. In Figure~\ref{fig:remeshing}, we find that for the analytic Wobbly Torus, \name reconstructs well when the proxy is reasonably close to the target and has at least 150 vertices, but with 100 vertices or fewer it begins to smooth over or miss high-detail features (see zoom-ins). This is expected: low-degree polynomials (quadratic here) have limited expressivity, and capturing small-scale detail typically requires either more parameters or a finer proxy; very coarse proxies also provide a weaker initialization and are more prone to local minima.

\begin{figure}[h!]
\includegraphics[width=\columnwidth]{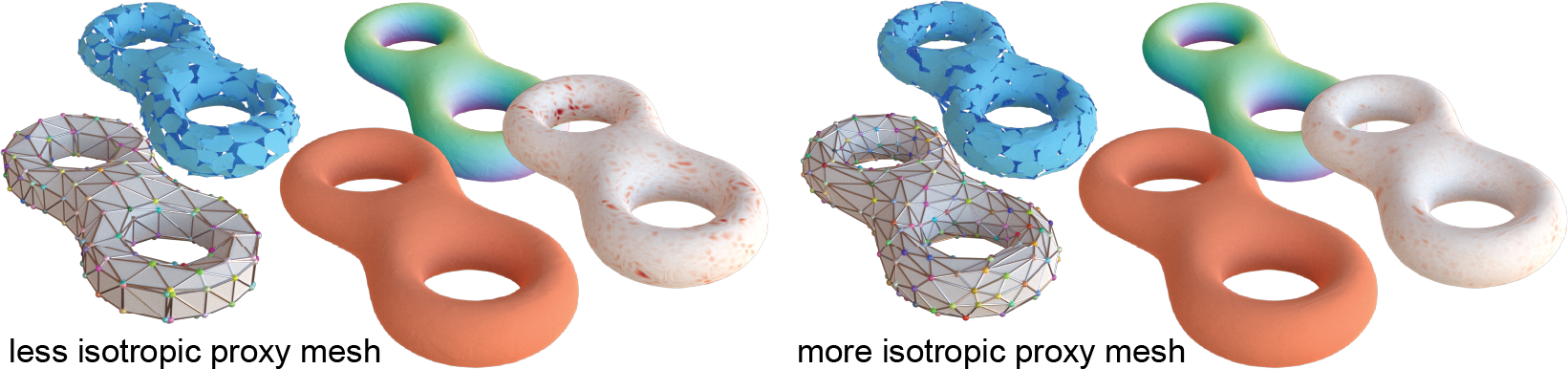}
\caption{\update{\textbf{Triangle quality.} We compare the BCS results on the same implicit surface (double torus) using proxy meshes with a similar number of faces but varying triangle quality. On the left, the proxy has a `less isotropic' triangulation with 436 triangles, produced directly by the marching cubes algorithm on the neural implicit function. On the right, the `more isotropic' triangulation with 500 triangles was produced by performing quadric edge collapse decimation on a higher resolution marching-cubes mesh. The reconstructed surface appears less wrinkled or creased on the BCS with the more isotropic proxy an the error is smaller.}}
\label{fig:triangle-quality}
\end{figure}

\begin{figure*}[t!]
    \centering
    \includegraphics[width=\linewidth]{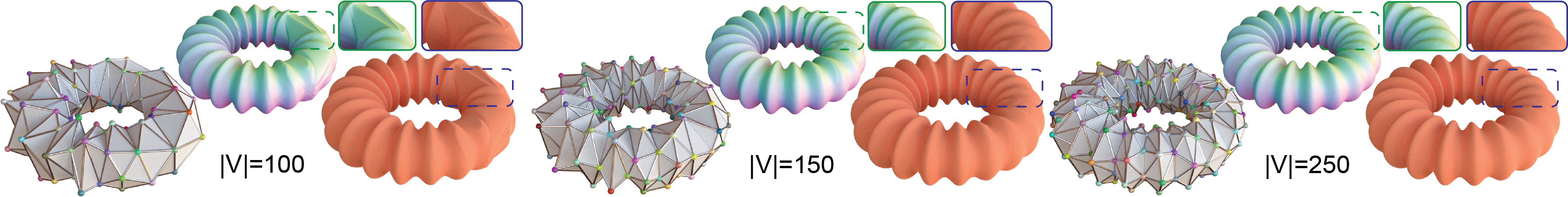}
    \caption{\textbf{Effect of the coarse proxy mesh on fitting quality.} We fit \name to a `rippled' target with strongly varying curvature using proxy meshes of increasing resolution ($|V|{=}100, 150, 250$). The results are largely robust to changes in proxy connectivity, reflecting the locality of our one-ring coordinates and PoU blending; however, at very low resolution, the local optimization can stall, leading to underfit regions in high-curvature areas (highlighted in the zoomed insets). Increasing proxy resolution improves fidelity in these challenging regions. We show both \name (front row) as well as corresponding zooms and normal maps (back row). }
    \label{fig:remeshing}
\end{figure*}

\update{In \Cref{fig:triangle-quality}, we investigate the effect of triangle quality in the proxy mesh, by comparing results on the same implicit surface (the double torus) using  proxy meshes with a similar number of faces (436 faces on the left, and 500 faces on the right). The proxy with a `low quality', `less isotropic' triangulation was produced directly by marching cubes algorithm on the neural implicit. The `high quality', `more isotropic' triangulation was produced by performing quadric edge collapse decimation on a higher resolution marching-cubes mesh. We see that the reconstructed surface appears more wrinkled or creased on the less isotropic BCS, and some of the higher-error regions correspond to visbly `skinny' triangles in the proxy. This justifies our decision to produce proxy meshes using quadric edge decimation on a high-resolution marching-cubes mesh, rather than directly use a lower-resolution marching-cubes mesh. }

\begin{figure}[b!]
    \includegraphics[width=\columnwidth]{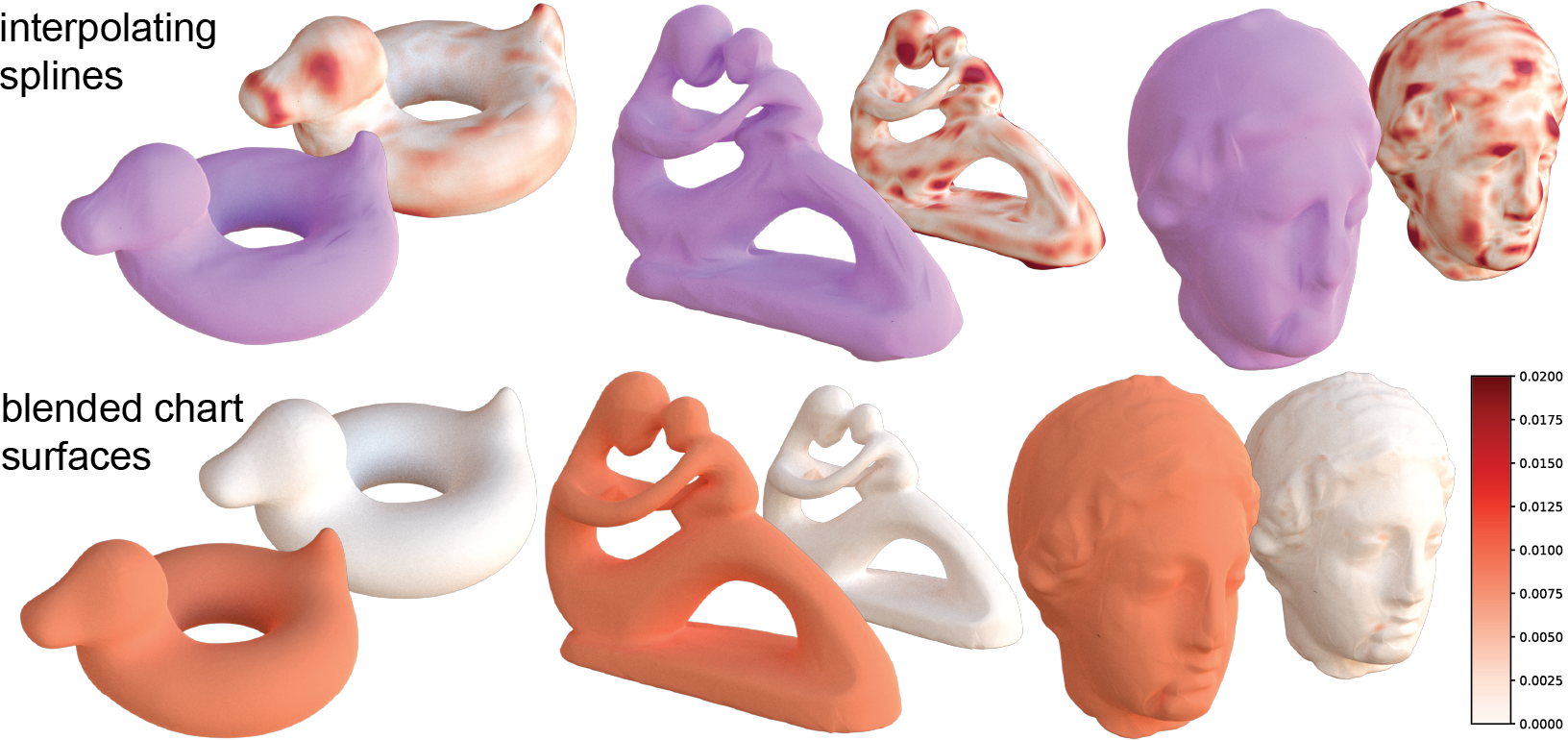}
    \caption{\textbf{Comparison} with Djuren et al. \cite{interpolatingSplines}. Whilst our Blended Chart Surface are constructed similarly to the Vertex-Centric Interpolating Splines (by blending together vertex functions, in this case polynomials), the key difference is that we jointly optimize vertex functions, guided by a target implicit shape, while Vertex-Centric Interpolating Splines use individually optimal vertex functions to best interpolate the coarse vertices. Error maps computed from the UDF show that ours achieves a more faithful surface reconstruction quality on the same coarse mesh, with the same type of vertex functions (quadratic polynomials).  }
    \label{fig:Djuren-comparison}
\end{figure}

\begin{figure}[t!]
    \centering
\includegraphics[width=\columnwidth]{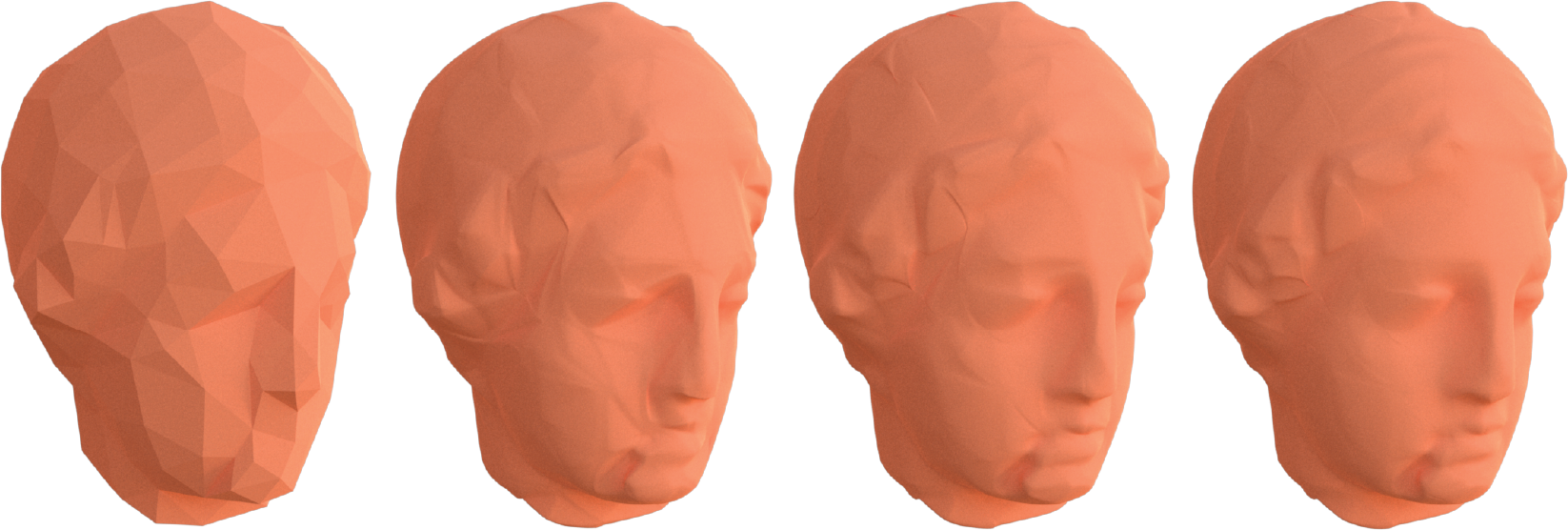}
    \caption{\textbf{Levels of detail controlled by polynomial degree.} We fit a Blended Chart Surface using cubic vertex polynomials and visualize coarser models by truncating coefficients to obtain constant (degree 0), linear (degree 1), and quadratic (degree 2) variants, respectively. Increasing degree improves local approximation quality.
    \update{From left to right, symmetric distance measured were 2.73e-3, 1.16e-3, 7.80e-4, 3.10e-4, normalized to a bounding box with a unit-length diagonal.}
    }
    \label{fig:truncation}
\end{figure}

\begin{figure}[t!]
    \centering
\includegraphics[width=\columnwidth]{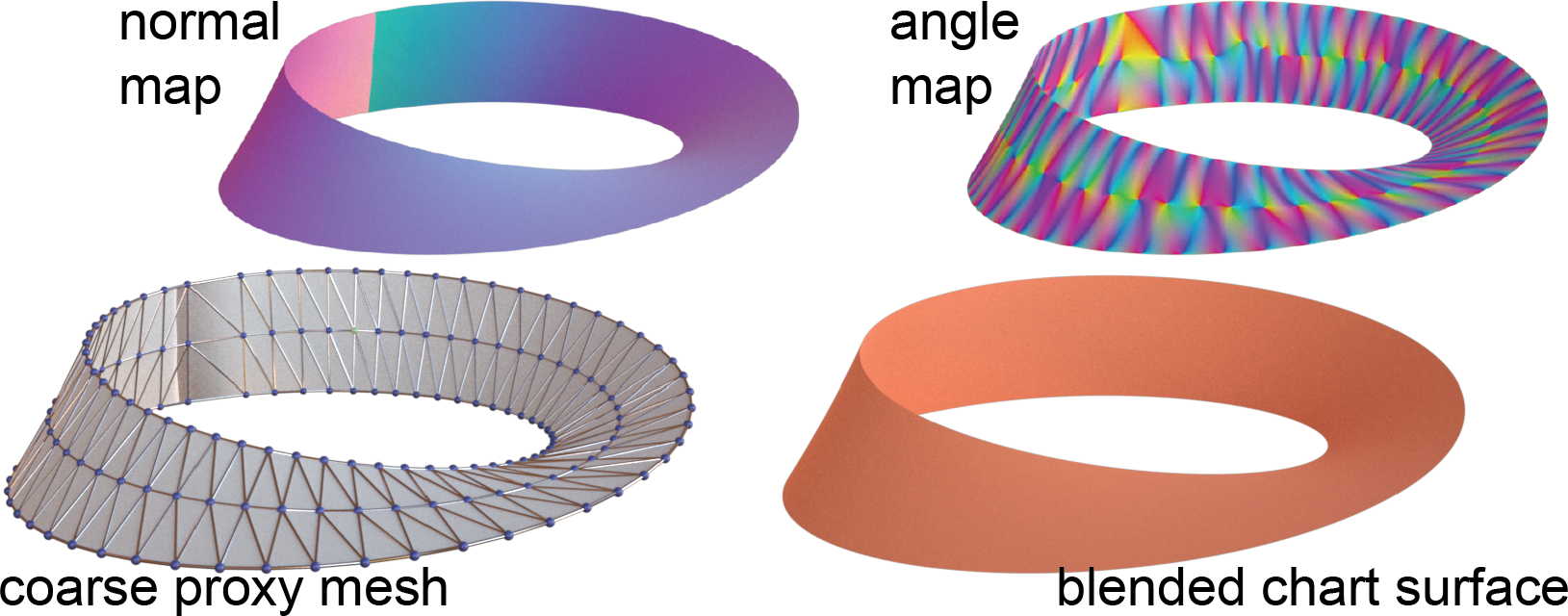}
    \caption{
    \textbf{Boundary handling and non-orientable support.} We reconstruct a M\"obius strip from a coarse proxy mesh, illustrating handling of boundaries and non-orientable connectivity. Since the method optimizes coupled local maps rather than a global parametrization, it extends naturally to non-orientable surfaces. The angles used in the one-ring coordinates are not oriented consistently (angles around each vertex are visualised using the HSV colourmap, and interpolated between vertices) but inconsistent orientation has no effect on the continuity guarantees. We use UDFs for both the M\"obius surface and its boundary, to define the reconstruction objective in this example, and trigonometric blending.}
    \label{fig:mobius}
\end{figure}

\paragraph*{Comparison to `Interpolating Splines'.}
In Figure \ref{fig:Djuren-comparison}, we compare the surface fitting quality of Blended Chart Surfaces to the most similar method, which is the Interpolating Splines of Djuren et al. \cite{interpolatingSplines}, with quadratic polynomials. We apply both methods using the same proxy meshes for Bob, Fertility and Igea (proxy meshes are the same as those shown in Figures \ref{fig:teaser} and \ref{fig:results}). The error-maps display the absolute value of the implicit function at each point (an approximate UDF). The `Interpolating Splines' unsurprisingly have high UDF in several areas because the method is only aware of the proxy mesh, with no other surface information.  Blended Chart Surfaces are qualitatively better fitted to the underlying surface, and quantitatively  achieve low UDF at surface points.

\paragraph*{Effect of Polynomial Degree.}
\name use polynomial vertex functions, naturally yielding a level-of-detail hierarchy. In \Cref{fig:truncation}, we optimize Igea with degree-3 (cubic) polynomials (7,560 coefficients) and visualize reconstructions after truncating to quadratic (4,536), linear (2,268), and constant-only (756) coefficients. (If proxy vertices lie exactly on the target surface, the constant terms can be omitted.) 
\update{The symmetric distance measure between the ground-truth surface ($S_1$) and the BCS ($S_2$) is the average of the orthogonal point-to-surface distance for two sets of points ($S(S_1))$ and $S(S_2))$) sampled on $S_1$ and $S_2$ respectively:
\begin{equation}
D(S_1, S_2) = \frac{1}{2|S(S_1)|} \sum_{p \in S(S_1) }d(p,S_2)
+
 \frac{1}{2|S(S_2)|} \sum_{p \in S(S_2) } d(p,S_1).
\end{equation}
We sampled 100,000 points uniformly on each surface and the meshes are scaled so that the bounding box of the ground-truth mesh has a unit-length diagonal. 
}

Constant-only coefficients are insufficient and produce visible sharpness, even though the underlying parametrization remains smooth (see Supplemental Section~B.7). Linear terms provide a large improvement, with further gains from quadratic terms; the jump from quadratic to cubic is subtler, making quadratic a favorable accuracy–compression trade-off in many settings. Apart from the truncation demonstration in \Cref{fig:truncation}, all BCS examples in this paper used quadratic polynomials.



\paragraph*{Equivariance Property.}
Blended Chart Curves and Blended Chart Surfaces are equivariant to rigid motions and to scaling, by construction (see Supplemental, sections B.1 and B.2). Equivariance and local control are desirable for animation and modeling applications.

\paragraph*{Boundaries and Non-Orientable Surfaces.}
The Blended Chart Surface framework is able to support surfaces with boundaries, and the boundary remains smooth if we construct the one-ring coordinates to be within a range of $\pi$, rather than $2\pi$; for boundary vertices  our `flattening' equation becomes 
$
(r,\theta) \mapsto \bigl(r\cos(3\theta/\eta),\; r\sin(3\theta/\eta)\bigr) $.

To ensure that the boundary of the domain maps to the intended curve, we include an additional loss on just the boundary edges:

\begin{equation}
\mathcal{L}_\text{boundary} = \frac{1}{|E_\text{boundary}|}\sum_{ij \in E_\text{boundary}} \int_{\mathbf{x} \in \text{edge}(ij)} \textbf{udf}(c_{ij}(\mathbf{x})) d \mathbf{x}
\end{equation}

Perhaps surprisingly, \name do not require orientability: blending is unchanged even if the angle $\theta$ is measured with inconsistent orientation at different vertices; one-ring coordinate orientations can be assigned arbitrarily. Figure~\ref{fig:mobius} shows a non-orientable surface with boundary (a Möbius strip), visualizing normals (with the necessary flip) and the per-vertex angle $\theta$ field (HSV, blended with our weights). The angle and normal colormaps reveal the orientation change, yet the Blended Chart Surface remains smooth. Since we cannot define signed distance function in this case, we optimize using analytically defined unsigned distance functions (UDFs) for both the interior surface and the boundary curve.


\begin{table}[b!]
    \centering
    
    \update{
    \caption{We quantify one aspect of surface quality by computing the extent of the BCS surface-area that is flipped in orientation with respect to the ground-truth surface (the original mesh).}

    \begin{tabular}{l|c}
        \hline
         & \textbf{\% Flipped Area} \\
        \hline
        \textit{\textbf{Neural SDFs}} & \\
        \hline
        Igea (500 faces), Degree 3 & 0.10 \\
        Igea (500 faces), Degree 2 (truncated) & 0.07 \\
        Igea (500 faces), Degree 1 (truncated) & 0.01 \\
        Igea (500 faces), Degree 2 & 0.15 \\
        
        Bob (500 faces), Degree 2 & 0.01 \\

        Fertility (500 faces), Degree 2 & 0.02 \\
        \hline
        \textit{\textbf{Analytic Implicits}} & \\
        \hline
        Wobbly Torus (500 faces), Degree 2 & 0.00 \\
        Wobbly Torus (300 faces), Degree 2 & 0.00 \\
        Wobbly Torus (200 faces), Degree 2 & 0.09 \\
        Twisted Torus (500 faces), Degree 2 & 0.01 \\
        Urchin (210 faces), Degree 2 & 0.00 \\

        \hline
    \end{tabular}
    
    }
    \label{tab:two_column_table}
\end{table}

\paragraph*{Compression and Storage Analysis.}
Our storage is dominated by polynomial coefficients (the number of polynomial coefficients is $3|V|   \binom{d_{\mathrm{poly}}+2}{2} \in \mathcal{O}(d_\text{poly}^2 |V|)$). For quadratic polynomials, that is $18 |V|$. For  most of the examples in this paper, the proxy mesh has around 250 vertices, so our representation stores $18 \times 250 = 4500 $ scalar coefficients (on top of the coarse mesh itself). The tiny MLP that is used in Figure \ref{fig:motivation} uses a similar number of parameters (4611, including the biases). Displacement MLPs that are used in practice, such as the one used in NGFM, have a larger number of parameters. From the network architecture described in the paper, we compute the number of parameters to be around 9000. Therefore our approach may offer a compression advantage.

\update{

\paragraph*{Measure of Flipped Normals.}
Because our fitting approach simply optimizes for the implicit function value to be close to zero, it is possible for `fold'/`wrinkle' artefacts to occur. In practice these artefacts tend to appear only on very small areas of the surface. We quantify this by meshing the BCS by pushing very densely meshed domain triangles, with 1024 sub-triangles each, through the parametrization and computing the proportion of these triangles (weighted by area) that have normals flipped with respect to the outwards-pointing normals of the ground-truth surface. For all examples analysed, the proportion  of flipped area is at most $0.15\%$.

}

\update{
\paragraph*{Sharp Features.}{

A limitation caused by guaranteed smoothness is that when the ground truth contains intentional sharp edges, these are over-smoothed in the BCS. In Figure \ref{fig:sharp-edges} we show five examples to demonstrate this limitation: the Oloid, the Fandisk, the Dice, the Arm, and the Gear. The results are over-smoothed, although the edges are still clearly defined as regions of high curvature. This is to be expected, because the BCS parametrization is smooth by construction, so it cannot represent sharp features unless the Jacobian determinant is exactly zero. It should also be noted that over-smoothing occurs when we generate the neural implicit representation from the ground-truth mesh; in these experiments a SoftPlus activation was used, which is a likely second cause of over-smoothing.

}
}

\update{

\begin{figure}[h!]
    \centering
\includegraphics[width=\columnwidth]{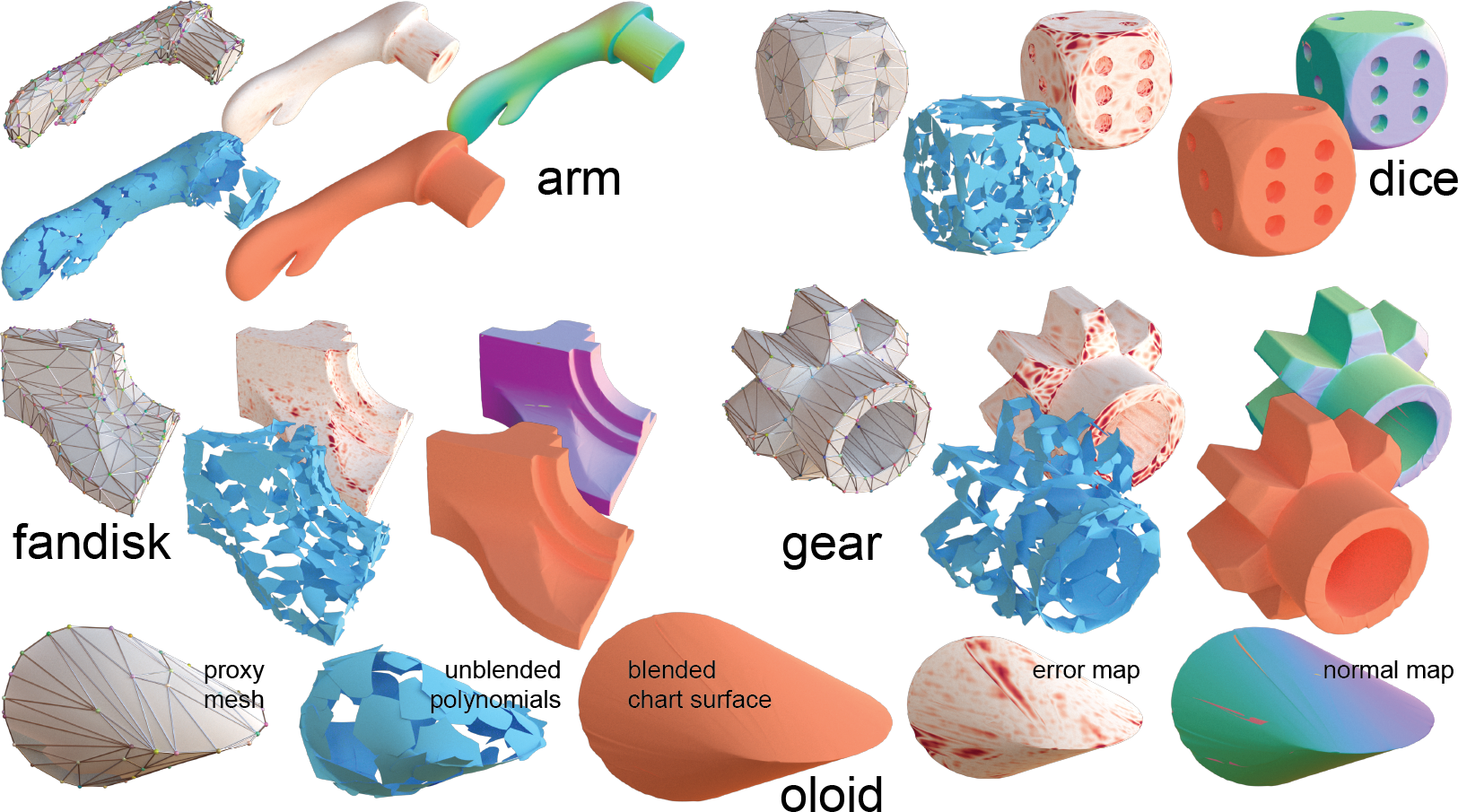}
    \caption{
    \update{
    \textbf{Limitations.} We show the results of BCS on five surfaces that contain sharp edges. The reconstructions are slightly over-smoothed and the highest errors generally occur along the intended sharp edge. This is as expected, because the BCS enforces a parametrisation that is mathematically smooth.}
    }
    \label{fig:sharp-edges}
\end{figure}

}

\section{Conclusion}


We introduced \name, a compact, explicit, and network-free surface representation that builds globally smooth geometry on top of a coarse simplicial proxy by optimizing per-vertex polynomial coefficients. Our key ingredients—one-ring coordinates and partition-of-unity blending—merge overlapping local maps into a single coherent surface with \textit{guaranteed smoothness} despite the proxy’s sharp edges, decoupling topology (from the proxy) from geometry (from local patches). The resulting model is fully differentiable, well-conditioned for optimization, and provides stable access to differential quantities such as normals and surface energies.

\name inherits the proxy topology and is sensitive to proxy quality: very coarse or poorly shaped proxies can hinder fitting and may require refinement or higher polynomial degree. Because local maps are not globally constrained to be injective, foldovers or self-intersections can occur without appropriate regularization, and the enforced global smoothness tends to round off genuinely non-smooth features such as creases. These limitations point to natural next steps: feature-aware, piecewise-smooth extensions with controlled continuity drops. As shown in \cite{interpolatingSplines}, vertex functions do not need to share the same class at each vertex and this may be used advantageously to improve surface-fitting e.g., by using spherical vertex functions in near-spherical areas, and constant vertex functions to produce sharp creases. Further interesting directions include time-varying formulations for animated surfaces via temporal regularization, and generative models that, conditioned on a coarse simplicial proxy and text, directly predict per-vertex coefficients to produce explicit surfaces in one shot.

\paragraph*{Acknowledgements}
The authors thank Navami Kairanda for valuable discussions on an earlier version of this work, Mariusz Tang and Anthony Steed for proofreading, and Yilin Liu for his patient technical help with the neural SDFs.
RW was
supported by the Engineering and Physical Sciences Research Council (grant number EP/S021566/1).




\renewcommand{\thesection}{\Alph{section}}


\setcounter{figure}{0}
\setcounter{table}{0}
\setcounter{equation}{0}
\setcounter{section}{0}

\renewcommand{\thefigure}{Supp.\arabic{figure}}
\renewcommand{\theequation}{Supp.\arabic{equation}}

\setcounter{secnumdepth}{2}

\title{
    Blended Chart Surfaces: A Seamless Explicit Representation for Smooth Surface Fitting\\
    \vspace{0.5em}
    \large\textit{Supplemental Material}
}
\date{}
\maketitle

\section{Precomputation}
During optimization, the tensor $A$ (which stores the polynomial coefficients of each vertex function) is the only quantity that is updated. Hence, we obtain substantial speedups by precomputing quantities that remain fixed. Before the optimization loop, we generate N samples per triangle of the coarse mesh and store them as a $N \times F \times 2$ tensor $\mathbf{x}$. The 2D samples are drawn uniformly at random on the canonical equilateral triangle. On these samples, we precompute the one-ring coordinates, blend weights, polynomial basis (i.e., $\Phi(\texttt{oc}(\mathbf{x}))$), rotations, radii, and angles.

During training, we draw batches of points per face from this precomputed pool. This substantially reduces the time required to evaluate the Blended Chart Surface output points prior to updating the polynomial coefficients. 
If the Jacobian of the parametrization is needed (e.g., for area-weighted sampling, normal regularization, or distortion regularization), we additionally precompute gradients to reduce the amount of \texttt{autograd} computation required in the training loop.
The Jacobian of the parametrization is,
\begin{eqnarray}
    \frac{ \partial c }{ \partial \mathbf{x} } &=& \frac{\partial c}{\partial \; \texttt{oc}(\mathbf{x})} \boxed{\frac{d  \;   \texttt{oc}(\mathbf{x})      }{d \mathbf{x}} }\; + \; \frac{\partial c}{\partial \; B^{2D}(\mathbf{x})}           \boxed{\frac{d \; B^{2D} (\mathbf{x})}{d \; \mathbf{x}}} \; \nonumber
    \\ &&+ \; \frac{\partial c}{\partial \; \Phi} \boxed{\frac{{d \; \Phi }}{d \; \texttt{oc}(\mathbf{x})}     }
    \boxed{\frac{d  \;   \texttt{oc}(\mathbf{x})      }{d \mathbf{x}} }
\end{eqnarray}
All quantities should be understood as tensors stacked over the appropriate indices (face indices and/or per-face vertex indices).
Note that the boxed terms do not depend on the polynomial coefficients and can therefore be precomputed. All boxed terms are obtained via \texttt{autograd} except for the polynomial-basis term $\frac{d\Phi}{d\texttt{oc}(\mathbf{x})}$. While this term could also be computed with \texttt{autograd}, it is simpler to exploit the closed-form differentiability of polynomials. If $\texttt{oc}(\mathbf{x})=(u,v)$, then for quadratic vertex functions (as used in the surface examples),


\begin{equation}
    \begin{aligned}
        \frac{d \Phi}{d \texttt{oc}(\mathbf{x})} &= \frac{d}{d (u,v)} 
        \begin{bmatrix} 1 & u & v & u^2 & uv & v^2 \end{bmatrix} \\
        &= \begin{bmatrix}
        0 & 1 & 0 & 2u & v & 0  \\
        0 & 0 & 1 & 0 & u & 2v 
        \end{bmatrix}.
    \end{aligned}
\end{equation}

\section{Proofs and Technical Details}
\subsection{Equivariance to Rotation and Scaling}

For both Blended Chart Curves and Blended Chart Surfaces (which we will jointly refer to as Blended Chart Shapes), the local geometry is defined by vertex functions $p_i(t) = s_i R_i m_i (t) + v_i$, which are blended together via

\begin{equation}
         c_{i,i+1}(t) = p_i(t)\,B(t) 
         + p_{i+1}(\tau_{i,i+1}(t))\,B(\tau_{i,i+1}(t)) \; \text{for} \; t \in (0,1].
     \label{eq:edge-c}
\end{equation}
(for curves) or

\begin{equation}
     c_{ ijk} (\mathbf{x}) =
     \begin{matrix}
         p_i(\texttt{oc}_i(\mathbf{x}))B^{\text{2D}}(\mathbf{x}, i)  
         +   p_j(\texttt{oc}_j(\mathbf{x}))B^{\text{2D}}(\mathbf{x}, j) 
          \\ + p_k(\texttt{oc}_k(\mathbf{x}))B^{\text{2D}}(
         \mathbf{x}, k)
     \end{matrix} 
     \label{eq:consistent_triangle_map}
\end{equation}
(for surfaces).

For the Blended Chart Shape to be equivariant to scaling, it means that if we scale the proxy (mesh) geometry by $s$, then the function $c$ will also be scaled by a factor of $s$.

Suppose we scale the proxy mesh geometry by $s$. Then $v_i$ scales by $s$, and $s_i$ also scales by $s$ because we defined the local scale $s_i$ to be the mean of the outgoing edge-lengths.
So clearly, $p_i$ also scales by $s$. Then we see that every term in the definition of $c_i$ scales by $s$. So the Blended Chart Shape is indeed equivariant to scaling.

Similarly, if we rotate the proxy geometry by $R$, then $v_i \mapsto Rv_i$, and also $R_i \mapsto R R_i$ because $R_i$ was defined to be a local coordinate frame based on the proxy mesh geometry. So overall, $p_i(t) \mapsto Rp_i(t)$. Every term in the definition of $c$ becomes rotated by $R$, therefore $c \mapsto R c$, and $c$ is equivariant to rotations of the proxy mesh geometry, as we wanted.

The key part of the formulation that allows equivariance to scale and rotation is that the vertex functions $p_i$ are constructed in a local coordinate frames $R_i$ based on the proxy geometry, and $p_i$ is proportional to a `local scale' term $s_i$.

\subsection{Equivariance to Translation, Partition of Unity}
A Blended Chart Shape (curve or surface) is equivariant to translations of the coarse points if and only if the Blending Function satisfies the Partition of Unity~(PoU) property.
\newline
\newline
\textbf{Proof for Blended Chart Curves:}
\newline
A blended chart curve is locally defined by the equation
\begin{equation}
         c_{i,i+1}(t) = p_i(t)\,B(t) 
         + p_{i+1}(\tau_{i,i+1}(t))\,B(\tau_{i,i+1}(t)) \; \text{for} \; t \in (0,1],
     \label{eq:edge-c}
\end{equation}
with 
$p_i(t) = s_i R_i m_i (t) + v_i$.

Now let
\begin{equation*}
  \hat{c}_{i,i+1}(t) =  
\hat{p}_i(t)B(t) 
         + \hat{p}_{i+1}(\tau_{i,i+1}(t))B(\tau_{i,i+1}(t)) \; \text{for} \; t \in (0,1]
\end{equation*}
for 
$\hat{p}_i(t) = s_i R_i m_i (t) + (v_i + T )$ where $T$ is some (nonzero) translation applied to all the coarse points. i.e, $\hat{p}_i (t) = p_i(t) + T$.

\noindent The blended chart curve is (locally) equivariant to translations of the coarse points 
$ \Leftrightarrow $
\newline
$\hat{c}_{i,i+1} (t) = c_{i,i+1} (t) + T$ for $t \in (0,1]$\\
$ \Leftrightarrow $

\begin{eqnarray*} 
    \hat{p}_i(t)B(t) 
         + \hat{p}_{i+1}(\tau_{i,i+1}(t))B(\tau_{i,i+1}(t))  = \\ 
         p_i(t)B(t) 
         + p_{i+1}(\tau_{i,i+1}(t))B(\tau_{i,i+1}(t)) + T \quad
 \text{for} \; t \in (0,1] 
\end{eqnarray*}
$ \Leftrightarrow $

\begin{eqnarray*}
    (p_i(t)+T)B(t) 
         + ({p}_{i+1}(\tau_{i,i+1}(t) + T))B(\tau_{i,i+1}(t))  = \\
         p_i(t)B(t) 
         + p_{i+1}(\tau_{i,i+1}(t))B(\tau_{i,i+1}(t)) + T \quad
          \text{for} \; t \in (0,1]
\end{eqnarray*}

$ \Leftrightarrow $
\begin{equation*}
     B(t) T
         +  B(\tau_{i,i+1}(t)) T = 
         T \quad
          \text{for} \; t \in [-1,0]
\end{equation*}
(by subtracting the common terms on both sides), 

Factoring out $T$, and using the assumption that $T \neq 0$, we see that the equation holds if and only if

\begin{equation*}
     B(t)
         +  B(\tau_{i,i+1}(t)) = 
         1 \quad 
          \text{for} \; t \in (0,1]\\
\end{equation*}

\noindent This is exactly the PoU property. So, equivariance to translation holds if and only if the blending function has PoU.

The same argument may be followed to show that Blended Chart Surfaces are equivariant to translation if and only if the 2D Blending functions satisfy PoU.

\textbf{Proof for Surfaces:}
The proof for Blended Chart Surfaces is completely analogous.
Define $\hat{p}_i$ in the same way as above, for an arbitrary non-zero translation $T$.

The BCS is locally defined by

\begin{equation}
     c_{ ijk} (\mathbf{x}) =
     \begin{matrix}
         p_i(\texttt{oc}^{ijk}_i(\mathbf{x}))B^{\text{2D}}(\mathbf{x}, i)  
         +   p_j(\texttt{oc}^{ijk}_j(\mathbf{x}))B^{\text{2D}}(\mathbf{x}, j) 
          \\ + p_k(\texttt{oc}^{ijk}_k(\mathbf{x}))B^{\text{2D}}(
         \mathbf{x}, k)
     \end{matrix},
\end{equation}
for $\mathbf{x}$ in the unit-length equilateral triangle which we will denote by $E$.

Let \begin{equation}
     \hat{c}_{ ijk} (\mathbf{x}) =
     \begin{matrix}
         \hat{p}_i(\texttt{oc}^{ijk}_i(\mathbf{x}))B^{\text{2D}}(\mathbf{x}, i)  
         +   \hat{p}_j(\texttt{oc}^{ijk}_j(\mathbf{x}))B^{\text{2D}}(\mathbf{x}, j) 
          \\ + \hat{p}_k(\texttt{oc}^{ijk}_k(\mathbf{x}))B^{\text{2D}}(
         \mathbf{x}, k)
     \end{matrix} 
\end{equation}

Abusing notation, we will drop the superscripts on the terms of the form $(\texttt{oc}^{ijk}_i(\mathbf{x}))$.

Blended Chart Surfaces are equivariant to translation of the coarse geometry if and only if 

$\hat{c}_{ijk} (\mathbf{x}) = c_{ijk} (\mathbf{x}) + T$ for $\mathbf{x} \in E$.\\
$ \Leftrightarrow $

\begin{eqnarray*} \hat{p}_i(\texttt{oc}_i(\mathbf{x}))B^{2D}(\mathbf{x},i) + \hat{p}_j(\texttt{oc}_j(\mathbf{x}))B^{2D}(\mathbf{x},j)  + \\\hat{p}_k(\texttt{oc}_k(\mathbf{x}))B^{2D}(\mathbf{x},k) 
          = \\ 
         p_i(\texttt{oc}_i(\mathbf{x}))B^{2D}(\mathbf{x},i) + p_j(\texttt{oc}_j(\mathbf{x}))B^{2D}(\mathbf{x},j)  + \\p_k(\texttt{oc}_k(\mathbf{x}))B^{2D}(\mathbf{x},k) 
         + T, \quad
 \text{for} \; \mathbf{x} \in E 
\end{eqnarray*}
$ \Leftrightarrow $

\begin{eqnarray*}
    ( p_i(\texttt{oc}_i(\mathbf{x})) + T)B^{2D}(\mathbf{x},i) + ( p_j(\texttt{oc}_j(\mathbf{x})) + T) B^{2D}(\mathbf{x},j)  + \\ ( p_k(\texttt{oc}_k(\mathbf{x})) + T)B^{2D}(\mathbf{x},k) = \\
         p_i(\texttt{oc}_i(\mathbf{x}))B^{2D}(\mathbf{x},i) + p_j(\texttt{oc}_j(\mathbf{x}))B^{2D}(\mathbf{x},j)  + \\p_k(\texttt{oc}_k(\mathbf{x}))B^{2D}(\mathbf{x},k) 
         + T, \quad
          \text{for} \; \mathbf{x} \in E
\end{eqnarray*}

$ \Leftrightarrow $
\begin{equation*}
     ( B^{2D}(\mathbf{x},i) + B^{2D}(\mathbf{x},j) + B^{2D}(\mathbf{x},k) ) T = 
         T, \quad
          \text{for} \; \mathbf{x} \in E,
\end{equation*}
by subtracting the common terms on both sides.

Using the assumption that $T \neq 0$, we see that the equation holds if and only if

\begin{equation*}
      B^{2D}(\mathbf{x}, i) +
         B^{2D}(\mathbf{x}, j) + B^{2D}(\mathbf{x}, k)  = 
         1, \quad
          \text{for} \; \mathbf{x} \in E.\\
\end{equation*}

\noindent This is exactly the PoU property. So, equivariance to translation holds if and only if the blending function has PoU.

\subsection{The 1D Blending Function is $C^k$ continuous}

We want to show that the `smooth transition' blending function $B$, as defined in Section 3.1, is smooth (i.e., $C^k$ continuous for all $k$).\par

Reminder:
\begin{equation}
B(x) = g \left (   \frac{|x|-(1-\beta)/2)}{\beta}   \right )
\end{equation}
where
\begin{equation}
    g(x) =  \frac{f(1-x)}{f(x)+f(1-x)} 
\end{equation}
and
\begin{equation}
f(x) = 
\left\{\begin{matrix}
 \text{exp}(-\frac{1}{x}) \quad \text{for} \quad x>0\\
0 \quad \text{otherwise.} 
\end{matrix}\right. 
\end{equation}

We first show that $f$ is smooth, and then argue that this implies $g$ and $B$ are smooth. The function $g$ is known as a Smooth Transition Function because it transitions between the functions that are constantly one, and constantly zero, and it has $C^\infty$ smoothness. \par

\textit{Showing that $f$ is smooth:}

Clearly, $\frac{d}{d^k x} f(x) =0$ for $x<0$, for all $k$. Applying the chain rule and product rule, we see that for $x>0$, every derivative has the form
\begin{equation}
    \frac{d}{d^k x} f(x) = P_{k} (x) e^{-x^{-1}}
\end{equation}
where $P_0 (x) = 1$, $P_1(x)=x^{-2}$ and in general, 
\begin{equation}
    P_{k+1}(x) = P'_k (x) + x^{-2} P_k (x)
\end{equation}

Without actually solving the recurrence, we see (inductively) that $P_k (x)$ always takes the form
\begin{equation}
    P_k(x) = a_{1,k}x^{-{(k+1)}} + ... + a_{k, k}x^{-2k}
\end{equation} for some scalar coefficients $a_{i,j}$.\par
Let $t=1/x$. Then $f_k(x)= f_k(1/t) = P_k(t)e^{-t}$
Now $P_k(1/t)$ is a polynomial in $t$, and by Algebra of Limits, $f_k (1/t) \rightarrow 0$ as $t \rightarrow \infty$. Therefore $f_k (x) \rightarrow 0$ as $x \downarrow 0$.

So the right hand and left hand limits agree, therefore $f$ is smooth including at zero.

A similar proof of the smoothness of the function $f$ is given by \cite{Nestruev2003SmoothManifolds}.

\textit{Showing that $g$ is smooth:}

Since $f(x)$ is smooth, then $f(1-x)$ is smooth, and $f(x)+f(1-x)$ is smooth. We also know that $f(x)+f(1-x)$ is strictly positive for all $x$, so $\frac{1}{f(x)+f(1-x)}$ is smooth. Therefore, $g(x)$ is the product of two smooth functions of $x$, so $g(x)$ is smooth.\par

\textit{Showing that $B$ is smooth:}

The parameter $\beta$ is constrained to be greater than zero. Then $B_+(x) = g \left (   \frac{x-(1-\beta)/2)}{\beta}   \right ) $ and $B_-(x) = g \left (   \frac{ - x -(1-\beta)/2)}{\beta}   \right )$ are clearly smooth because they are simple transformations of $g(x)$.\par
It remains to show that the left hand and right hand limits agree. That is trivially true, because $B_+$ and $B_-$ are both constant (equal to 1) in a neighborhood of zero (because we set $\beta$ strictly less than 1), so all derivatives on both side are zero.\par
Therefore $B$ has $C^\infty$ continuity, as required.

\subsection{The Blended Chart Curve is $C^k$ Continuous}

For this part it is cleanest to extend $B$ and the $p_i$ to functions that are zero outside of $[-1,1]$. Then we can just write
\begin{equation*}
    c_i(t) = p_i(t)B(t) + p_{i-1}(t+1)B(t+1) + p_{i+1}(t-1)B(t-1)
\end{equation*} for $t \in [-1,1]$ and we remove the need for cases in the definition.\par

The General Leibniz Rule states that if $f$ and $g$ are both $n$-times differentiable, then their product $fg$ is $n$-times differentiable (see \cite{olver1993applications}, p. 318). The $n^\text{th}$ derivative may be expressed using binomial coefficients, but we only require the existence statement.\par

Now, all of the $p_i$ functions are $k$-times differentiable by assumption, so $p_i(t)$, $p_{i-1}(t+1)$ and $p_{i+1}(t-1)$ are all $k$-times differentiable (using the chain rule), and we showed above that $B(t)$ is $k$-times differentiable for all $k$, therefore $B(t-1)$ and $B(t+1)$ are also $k$-times differentiable. Importantly, $B$ is zero at $\pm 1$ and the first $k$ derivatives of $B$ are also zero at $\pm 1$ (Properties C2 and C3). This ensures that the products between local maps $p_i$ and blending functions $B$ are also $C^k$ on the \textit{extended} domain, so smoothness is not broken at the vertex locations (i.e. when $t=0$ or $t=\pm 1$).

All of the terms $p_i(t)B(t)$, $p_{i-1}(t+1)B(t+1)$ and $p_{i+1}(t-1)B(t-1)$ are $k$-times differentiable, and so is their sum $c_i(t)$ (by linearity of differentiation).

So $c_i$ is $C^k$ continuous. i.e. $c_i$ inherits the same level of smoothness as the $p_i$ functions, and as long as we construct the $p_i$ functions to be $C^\infty$ continuous then the $c_i$ is $C^\infty$ continuous.

\subsection{Allowable range of $\alpha$ , for Blended Chart Surfaces}
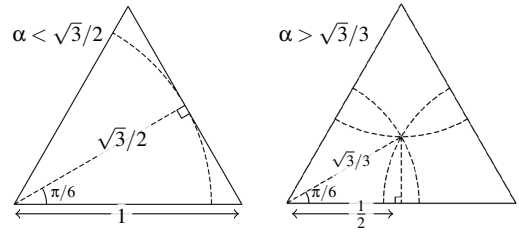
\begin{figure}[t!]
\centering
\begin{tikzpicture}
  \node[anchor=south west, inner sep=0] (img) 
    at (0,0) {\includegraphics[width=200pt]{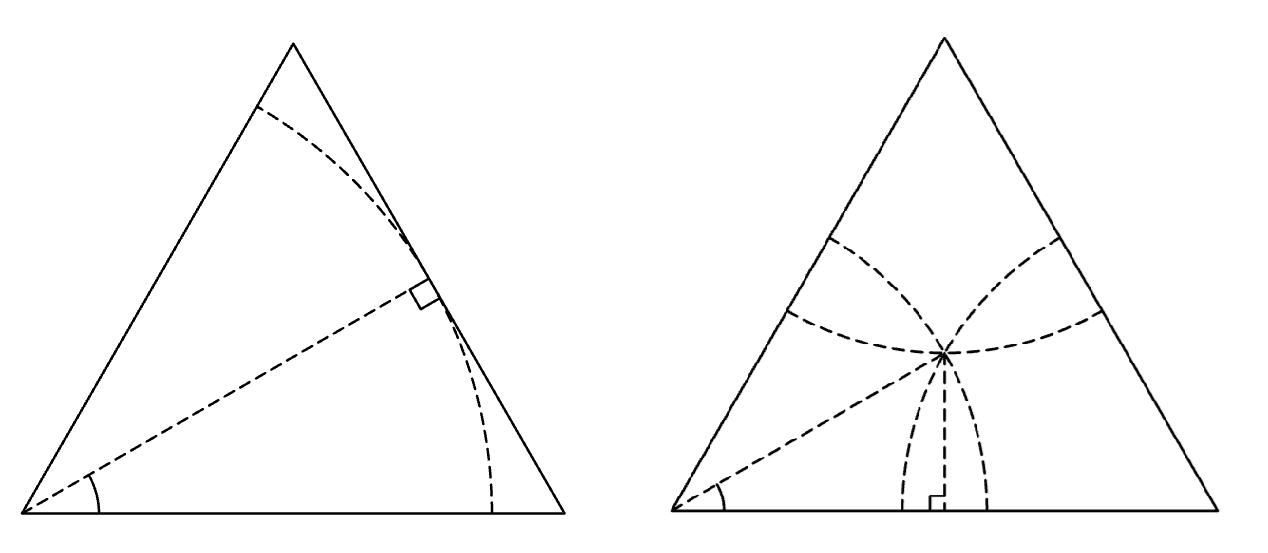}};
  \begin{scope}[x={(img.south east)},y={(img.north west)}]

    \node[
  fill=white,
  fill opacity=0.85,
  text opacity=1,
  inner sep=2pt,
  font=\footnotesize,
  rounded corners
]  at (0.22,0.35) {$\sqrt{3}/2$};

\node[
  fill=white,
  fill opacity=0.0,
  text opacity=1,
  inner sep=2pt,
  font=\tiny,
  rounded corners
]  at (0.11,0.12) {$\pi / 6$};

\node[
  fill=white,
  fill opacity=0.0,
  text opacity=1,
  inner sep=2pt,
  font=\footnotesize,
  rounded corners
]  at (0.1,0.8) {$\alpha < \sqrt{3}/2$};

    \node[
  fill=white,
  fill opacity=0.85,
  text opacity=1,
  inner sep=2pt,
  font=\tiny,
  rounded corners
]  at (0.65,0.26) {$\sqrt{3}/3$};

\node[
  fill=white,
  fill opacity=0.0,
  text opacity=1,
  inner sep=2pt,
  font=\tiny,
  rounded corners
]  at (0.6,0.12) {$\pi / 6$};

\node[
  fill=white,
  fill opacity=0.0,
  text opacity=1,
  inner sep=2pt,
  font=\footnotesize,
  rounded corners
]  at (0.6,0.8) {$\alpha > \sqrt{3}/3$};

\draw[->, thin] (0.22,0.03) -- (0.02,0.03);
\draw[->, thin] (0.22,0.03) -- (0.44,0.03);

    \node[
  fill=white,
  fill opacity=0.85,
  text opacity=1,
  inner sep=2pt,
  font=\footnotesize,
  rounded corners
]  at (0.22,0.02) {$1$};

    \draw[->, thin] (0.67,0.03) -- (0.54,0.03);
\draw[->, thin] (0.67,0.03) -- (0.73,0.03);

\node[
  fill=white,
  fill opacity=0.85,
  text opacity=1,
  inner sep=2pt,
  font=\footnotesize,
  rounded corners
]  at (0.67,0.03) {$\frac{1}{2}$};

  \end{scope}
\end{tikzpicture}

\caption{Allowable range of $\alpha$ parameter (the radius of the blending function's non-zero region). Left: $\alpha < \cos(\pi / 6)=\sqrt{3}/2$, so that the blending function reaches zero \textit{before} it reaches the opposite edge. Right: $\alpha > \frac{1}{2} / \cos(\pi/6) = \sqrt{3}/3$, so that there is a non-zero amount of overlap in the support of the blending functions from each vertex.}
\label{fig:alpha_figure}
\end{figure}

To ensure that each blending function reaches zero before reaching the opposite edge, $\alpha$ must be smaller than the distance to the opposite edge. (If the blending function does not reach zero before reaching the opposite edge, then there is a jump when the induced `hat function' drops to zero outside of the one-ring, and this leads to discontinuities/ holes in the Blended Chart Surface.) Since the domain triangle $E$ is equilateral with unit side-lengths, this means that $\alpha < \cos (\pi / 6) = \sqrt{3}/2$. (See  \Cref{fig:alpha_figure}, left.)\par

Additionally, in order for the sum of the blending functions to always be positive (so that it's possible to normalize to get PoU blending functions), we require that $\alpha$ is greater than the distance to the centre of the triangle. That is, $\alpha > \sqrt{3}/3$. (See  \Cref{fig:alpha_figure}, right.)
Overall, the allowable range of the `overlap parameter' $\alpha$ is
$\sqrt{3}/3 < \alpha < \sqrt{3}/2$, or approximately,
$0.577 < \alpha < 0.866$.

\subsection{The one-ring coordinates map and the transition maps are $C^\infty$ diffeomorphisms}


The `flattening' map that converts equilateral-triangle coordinates to one-ring coordinates is a $C^\infty$ diffeomorphism on the punctured one-ring. This is because, except at the origin, it can be expressed in polar coordinates as a smooth composition of three maps: the coordinate change $(x,y)\mapsto(r,\theta)$, the linear angular rescaling $(r,\theta)\mapsto\bigl(r,\frac{6 \theta}{\eta}\bigr)$, and the inverse coordinate change $(r,\phi)\mapsto(r\cos\phi,r\sin\phi)$. The composition of smooth maps is smooth. The Jacobian determinant is constant and equal to $\frac{6}{\eta}$ everywhere (which is non-zero); geometrically, the map scales the angular direction by $\tfrac{6}{\eta}$ while leaving the radial direction unchanged. Therefore the `flattening' map is a $C^\infty$ diffeomorphism.

The composition of an `unflattening' and `flattening' map is still a $C^\infty$ diffeomorphism as long as none of the points lies on one of the `centres of flattening' (because diffeomorphisms are closed under composition and inversion). Therefore the \textit{transition maps} that convert between one-ring coordinates centred on two different vertices, for points in the interior of the rhomboidal overlap region between two one-rings, are $C^\infty$ diffeomorphisms.

\subsection{The 2D Blending Function has $C^\infty$ continuity }
The $2D$ blending function is, by construction, $1$ at one vertex in the equilateral parametrization domain and then drops outward from the vertex, before reaching zero at the radius $\alpha$. With a value of $\alpha$ in the valid range, then the radially symmetric heightfield swept out by 
$B$ is smooth on the equilateral triangle domain (smoothness is ensured at the central vertex, because our choice of the parameter $\alpha$ ensures that $B$ is locally flat in a finite neighborhood of both one and zero).

A corollary of the General Leibniz Rule states that if $a$ and $b$ are both $n$-times differentiable, and $b$ is nowhere zero, then $a / b$ is $n$-times differentiable. (This follows because $a/b$ can be expressed as the product $(a)(1/b)$, and $1/b$ is $C^n$ because the reciprocal function is $C^\infty$ except at zero, and the composition of $C^n$ functions is also $C^n$.)

Applying this to the definition of $B^{2D}$, the numerator and denominator are both $C^\infty$ and since we took $\alpha$ to be greater than $\sqrt{3}/3$, then the denominator is nowhere zero, therefore $B^{2D}$ is $C^\infty$. In other words, when we normalize to ensure PoU, this doesn't break the smoothness of the 2D blending function.

We have shown that $B^{2D}$ is $C^\infty$ with respect to the equilateral triangle domain. In the next section we will need to use the fact that $B^{2D}$ is $C^\infty$ with respect to the \textit{one-ring coordinates}. 

\textit{Showing that $B^{2D}$ is $C^\infty$ with respect to one-ring coordinates in a single triangle:}

The flattening map (from equilateral one-rings to flat one-ring coordinates) is a $C^\infty$ diffeomorphism, everywhere except the central vertex (refer to section C.6). Therefore $B^{2D}$ is $C^\infty$ with respect to one-ring coordinates in a single triangle, except possibly at the central vertex. In fact, $B^{2D}$ is also $C^\infty$ at the central vertex, because our construction ensures that $B^{2D}$ is locally flat, equal to one, in a neighborhood of the vertex (this follows from properties S2 and S3), so the derivatives of all degrees are zero there.

\textit{Showing that $B^{2D}$ is $C^\infty$ with respect to one-ring coordinates in the whole one-ring:}

It isn't immediately obvious that the `hat function' formed by tiling together the blending functions on the $\eta$ triangles into a heightfield on a whole one-ring, is still $C^\infty$ along the edges (with respect to one-ring coordinates).

This `hat function' \textit{is} $C^\infty$ with respect to one-ring coordinates though, because $B^{2D}$ was constructed to have all directional derivatives equal to zero along the direction perpendicular to the edges, at points on the edges (property S3), so there is no discontinuity in the gradient or any of the higher derivatives as we cross over an edge.


\subsection{The Blended Chart Surface has $C^\infty$ Continuity}

We would like to show that the Blended Chart Surface has $C^\infty$ smoothness. To be more precise, we want to show that the Blended Chart Surface has $C^\infty$ smoothness with respect to some overlapping local charts. In particular, we take the local charts to be the one-ring coordinates on the interior of the one-rings (slightly inside the edge of the one-rings, to avoid the outer vertices).


Similar to the proof in the curves case, we can extend the blending functions with zeros outside of the one-ring, to allow us to express the overlapping local-maps $c_i$ without the need for cases:

\begin{equation}
    c_i (\texttt{oc}_i (\mathbf{x})) = B^{2D}(\mathbf{x},i)  p_i (\texttt{oc}_i (\mathbf{x})) + \sum _{j : (i,j) \in \text{edges} }B^{2D}(\mathbf{x},j)  p_j (\texttt{oc}_j (\mathbf{x}))
\end{equation}

Each of the terms has $C^\infty$ continuity with respect to its own one-ring coordinates; this follows from the $C^\infty$ smoothness of the blending function with respect to one-ring coordinates (section C.7), and the $C^\infty$ smoothness of the $p_i$ functions (which are rigidly transformed polynomials, in our examples).

In addition, the \textit{transition maps} that convert between pairs of one-ring coordinates, e.g. $\texttt{oc}_i(\mathbf{x})$ and $\texttt{oc}_j(\mathbf{x})$, are $C^\infty$ diffeomorphisms, away from the vertices $i$ and $j$ (as shown in section C.6).

This means that all of the terms are also $C^\infty$ smooth with respect to the one-ring coordinates of vertex $i$. Finally, the class of smooth functions is closed under addition, so $c_i$ is $C^\infty$ smooth with respect to the one-ring coordinates of vertex $i$.

\subsection{Can $c_i$ be non-injective, or have a zero-derivative?}
Unlike a genuine Atlas, which consists of `charts' from regions of the shape to $\mathbb{R}^d$, we work with \textit{local maps} the other way around: from regions of $\mathbb{R}^d$ to the shape.\par
This construction does not necessarily constrain $c_i$ to being injective. So, even though $c_i$ is $C^k$ continuous, the image of the curve may contain:
\begin{itemize}
    \item self-intersections
    \item singularities where $\frac{d}{dt} c_i (t) = 0$.
\end{itemize}
Curves that self-intersect are non-manifold. An example would be the path of a particle that moves smoothly and visits the same point at multiple points in time.
Singularities may lead to `cusps' in the trace of the curve. (Imagine the path of a particle that slows down to velocity zero, then changes direction when it resumes movement.)
For example, if we specifically initialize the vertex functions to be constant points then there are singularities, which show up as sharp corners on the trace of the initialized blended chart curve, but these corners disappear as soon as the $p_i$ functions are perturbed away from zero.

 Cusps are undesirable in our curve representation, but this is not practically an issue, because (although we don't prove this) the probability that the components of a $c_i$ function have zero-derivative at exactly the same point, simultaneously in $x$ and $y$ directions, is zero so this will never actually happen after the initialization.

 In the case of Blended Chart Surfaces, self-intersections may also occur when $c_i$ is non-injective. If $c_i$ is singular (some directional derivative is zero) then there can be `cusp'-like places, where the surface still has a local parametrisation but the parametrisation `slows down to zero'. For example, we see sharp edges if we set all of the vertex functions to be constant points (see Figure 10, far left). But in practice, the probability of the Jacobian being singular after initialisation is zero.


\subsection{Is there unique set of vertex functions for a given parametrized shape?}

We discuss in the curves case. The question is whether there is a one-to-one correspondence between ordered lists of vertex functions and parametrized curves. In other words, given all of the $c_i$ and knowledge of the blending function, hyperparameters $\alpha, s_\text{global}$ etc., can we recover all of the $p_i$?\par

When the $p_i$ belong to a very large class of functions, one can imagine perturbing the three vertex functions for a triangle in such a way that the final blended function, $c_i$, looks the same or similar. If the class of functions is much smaller, such as polynomials, we could take sample points of the $c_i$ and construct a linear system to solve for the polynomial coefficients. This may be an interesting direction for future work and has relevance to generative models (where it is advantageous for the latent space of `shape parameters' to be compact).

\bibliographystyle{eg-alpha-doi} 
\bibliography{bns_references}

\end{document}